\documentclass[journal]{IEEEtran}
\usepackage{ifpdf}
\usepackage{cite}
\usepackage{graphicx}
\usepackage{footnote}
\usepackage[ruled,vlined]{algorithm2e}
\usepackage{comment}
\usepackage{multirow}
\usepackage{amssymb}
\usepackage[hyphens]{url}
\usepackage{color,soul}
\usepackage{subfig}
\DeclareGraphicsExtensions{.eps}
\usepackage{amsmath}
\usepackage{threeparttable}
\usepackage{balance}
\usepackage{float}
\usepackage{soul}
\usepackage{makecell}
\usepackage{rotating}
\graphicspath{{images/}{../images/}}

\usepackage{lipsum}
\usepackage{mathtools}
\usepackage{cuted}

\begin{document}


\title{Contextual Intelligent Decisions: Expert Moderation of Machine Outputs for Fair Assessment of Commercial Driving}

\author{Jimiama Mafeni Mase $^{1}$, Direnc Pekaslan$^{1}$, Utkarsh Agrawal$^{2}$,  Mohammad Mesgarpour,$^{3}$\\
 Peter Chapman $^{4}$, Mercedes Torres Torres$^{1}$, Grazziela P. Figueredo$^{1}$\\ 
$^1$School of Computer Science, The University of Nottingham\\
$^2$School of Medicine, University of St Andrews\\
$^4$School of Psychology, The University of Nottingham\\
$^3$Microlise, Farrington Way, Eastwood, Nottingham}
\maketitle
\begin{abstract}
Commercial driving is a complex multifaceted task influenced by personal traits and external contextual factors, such as weather, traffic, road conditions, etc. Previous intelligent commercial driver-assessment systems do not consider these factors when analysing the impact of driving behaviours on road safety, potentially producing biased, inaccurate, and unfair assessments. In this paper, we introduce a methodology (Expert-centered Driver Assessment) towards a fairer automatic road safety assessment of drivers' behaviours, taking into consideration behaviours as a response to contextual factors. The contextual moderation embedded within the intelligent decision-making process is underpinned by expert input, comprising of a range of associated stakeholders in the industry. Guided by the literature and expert input, we identify critical factors affecting driving and develop an interval-valued response-format questionnaire to capture the uncertainty of the influence of factors and variance amongst experts' views. Questionnaire data are modelled and analysed using fuzzy sets, as they provide a suitable computational approach to be incorporated into decision-making systems with uncertainty. The methodology has allowed us to identify the factors that need to be considered when moderating driver sensor data, and to effectively capture experts' opinions about the effects of the factors. An example of our methodology using Heavy Goods Vehicles professionals input is provided to demonstrate how the expert-centred moderation can be embedded in intelligent driver assessment systems.

\end{abstract}


\section{Introduction}
\label{intro}
Most intelligent driver-assessment and profiling systems for road safety use individual sources of data, such as telematics~\cite{Figueredo2018, mase2020capturing,agrawal2019towards,constantinescu2010driving,Ellison2012}, video footage~\cite{mase2020hybrid,theagarajan2018deepdriver,xiao2019fatigue}, phone usage~\cite{johnson2011driving,Chen2015} and physiological signals~\cite{foy2018mental,yeo2009can}. Multiple sources of driver data used simultaneously have also been employed for more robust assessments of driver behaviours~\cite{carmona2015data,daza2011drowsiness}. Those studies however do not consider contextual characteristics of driving, such as drivers' physical and mental states, in-vehicle actions (e.g. operation of in-vehicle technologies), weather conditions, traffic conditions, road geometry, road types, work schedules, drivers' reactions to events, other vehicles, and road users. These factors impact drivers' responses and are mostly not captured in driver data~\cite{vilaca2017systematic,osafune2017analysis}. 

The influence of those contextual factors on driving should be utilised to moderate data-driven, road safety assessments of driver behaviours, thereby, producing fairer, explainable, and reliable intelligent driver behaviour assessment systems~\cite{arrieta2020explainable}. For example, in Figueredo \textit{et al}~\cite{figueredo2015data}, the authors use telematic incident data to identify top performing Heavy Goods Vehicle (HGV) drivers in the United Kingdom (UK) in terms of safe and economic driving; 'Microlise Driver of the Year Awards'. However, they do not consider the inevitable negative external factors (e.g. time pressure~\cite{mic2020} and poor weather conditions~\cite{hakkanen2001}) that could affect drivers during their journeys. This example illustrates the danger of unfairness and the need for human moderation of the endpoint data collected by assessing the circumstances affecting driving behaviours that may lead to road incidents. Such intelligent systems that consider context during processing and inference are crucial in ensuring trust, acceptance, and successful adoption among stakeholders~\cite{toreini2020relationship,vaughan2020human}. Furthermore, in the era of big data streams, the human moderation should be embedded into intelligent systems for automatic assessment as it is infeasible to regulate the data manually~\cite{gama2010knowledge}. 

Historically, questionnaires and surveys have been the main tools to capture the influence of contextual factors on driving behaviours~\cite{mahajan2019effects,foy2018mental,stern2019data,wu2018influence,kidd2016influence}. However, the approaches used in previous studies have the following limitations:
\begin{enumerate}
\item They do not include the input of stakeholders during the identification of factors and the design of data collection tools. This is essential to ensure that the factors identified are accurate and realistic to driving, and the questionnaires are understandable and easy to complete~\cite{mao2005state}; 
\item There is a bias towards the sole opinion of drivers, therefore neglecting other stakeholders in the industry who may possess important qualitative insights about the factors. In order to rely on subjective responses to improve decision making, we need an inclusive system that considers the viewpoints of all major stakeholders in the industry who are directed or indirectly affected by the system's outcomes~\cite{arrieta2020explainable};
\item The complexities of driving, different interpretations of questions, different levels of indecision, experience and expertise of participants increases the levels of uncertainty and variability regarding expert input. An adequate moderation system needs to account for this.
\item  They do not illustrate how the perceived effects of contextual factors could be embedded into intelligent driver systems.
\end{enumerate}


In this study, we present a methodology called \textbf{E}xpert-centered \textbf{D}river \textbf{A}ssessment (EDA) to identify, understand, and model the impact of contextual factors on the safety of commercial driving. Guided by the literature and with the help of stakeholders, we identify important factors that affect driving both positively and negatively. Workshops with domain experts validate the factors and the questionnaires designed. The questionnaires are later distributed to gather expert views on how those factors affect driving. Fuzzy Sets (FSs) based Interval Agreement Approach (IAA)~\cite{wagner2014interval} are used to model the variance between expert views~\cite{ellerby2020capturing,ellerby2019exploring}. The outcomes can then be incorporated into intelligent decision systems~\cite{mase2020capturing}. We demonstrate our methodology by investigating a case study for HGV driving in the UK.




\section{Background}
\label{background}

\subsection{Capturing the effects of contextual factors}
\label{CapturingContextualFactors}
Questionnaires and surveys have been the main tools to capture the influence of contextual factors on driving behaviours~\cite{mahajan2019effects,foy2018mental,stern2019data,wu2018influence,kidd2016influence}. The tools consist of discrete-valued response-format questions asking participants (a) to rate the extent to which they agree with statements relating to how different factors affect driving behaviours~\cite{foy2018mental} ; (b) how often the participants engage in risky driving ~\cite{mahajan2019effects,stern2019data,wu2018influence}; and/or (c) rank factors based on their contributions to accidents~\cite{kidd2016influence}. Examples of such discrete-valued response-format scales are a seven-point Likert scale (from 1, `strongly disagree', to 7, ‘strongly agree’) and a six-point scale (from 0, `never', to 5, `all the time') where participants are asked to select a single response. The responses captured using these tools are analysed using statistical techniques to identify statements with the highest agreement, identify significant predictors of incidents or accidents, causation and correlation amongst factors, e.g. Path analysis~\cite{poulter2008application}, p-value of means~\cite{hakkanen2001}, Logistic regression~\cite{sullman2002aberrant,hakkanen2001}, and Pearson Correlation~\cite{sullman2002aberrant,foy2018mental}. 

As mentioned in the Introduction Section, the above approaches are limited in: 1) their questionnaire design; 2) recruiting only drivers as participants; 3) they do not consider the uncertainties of the effects of contextual factors produced by complex relationships; 4) their analytical techniques do not provide a clear representation and separation of experts' opinions as knowledge varies among experts; and 5) lack of incorporation of contextual factors into intelligent driver systems. In this work, we address these gaps by presenting a methodology that: 1) Includes domain experts during the identification of important factors and the design of data collection tools to ensure that the factors are up-to-date, accurate and realistic of driving, and the questionnaire is understandable and easy to complete, 2) uses a wider cohort of stakeholders for a comprehensive provision of the influence of factors, 3) uses a recently proposed interval-valued response-format questionnaire~\cite{ellerby2019decsys} to capture the uncertainties in the effects of factors, and 4) uses fuzzy sets to model and aggregate experts' responses within the same profession and across different professions. These fuzzy sets provide a clear representation of the variability in the viewpoints of experts and can be integrated into intelligent systems to regulate driver-assessment.

\subsection{Variability and uncertainty in knowledge}
\label{uncertknowledge}

Information obtained from experts will likely contain uncertainty due to the complex relationships between contextual factors and driving performance. For example, the precise effect of driver-facing cameras on a driver’s performance may be difficult to determine as the relationship between driver-facing cameras and driving performance is not clear-cut due to interactive effects of other factors such as, the type of intervention after monitoring. In addition, knowledge varies even among experts within the same profession due to different interpretations of questions, ambiguity of questions, different levels of indecision, and different experiences~\cite{ellerby2019decsys}. Thus, it is intuitively expected that different experts ---even though they have similar roles--- may provide different answers to questions. When experts focus is changed, the differences in the viewpoints of different professions may also arise, due to their distinct responsibilities, roles and expectations~\cite{navarro2019measuring}. For example a group of researchers may focus on the aspect of scientific knowledge of road safety by using experiments, whereas fleet managers may focus on optimising the delivery of goods and services in their companies. Thus, different levels of uncertainty exist in knowledge captured from a wide cohort of stakeholders, which must be effectively modelled to provide a comprehensive, reliable and clear representation of knowledge for decision making~\cite{ellerby2020capturing,ellerby2019exploring}.

Recently, a fuzzy computational technique called Interval Agreement Approach (IAA)~\cite{wagner2014interval} is proposed to model and provide a clear representation of these uncertainties in experts' opinions by following a least commitment principle~\cite{hu2012least} to prevent any loss of information.

\subsection{Interval Agreement Approach}
\label{iaa}

\begin{figure}[!htpb]
\centering
\scalebox{0.33}{
\includegraphics{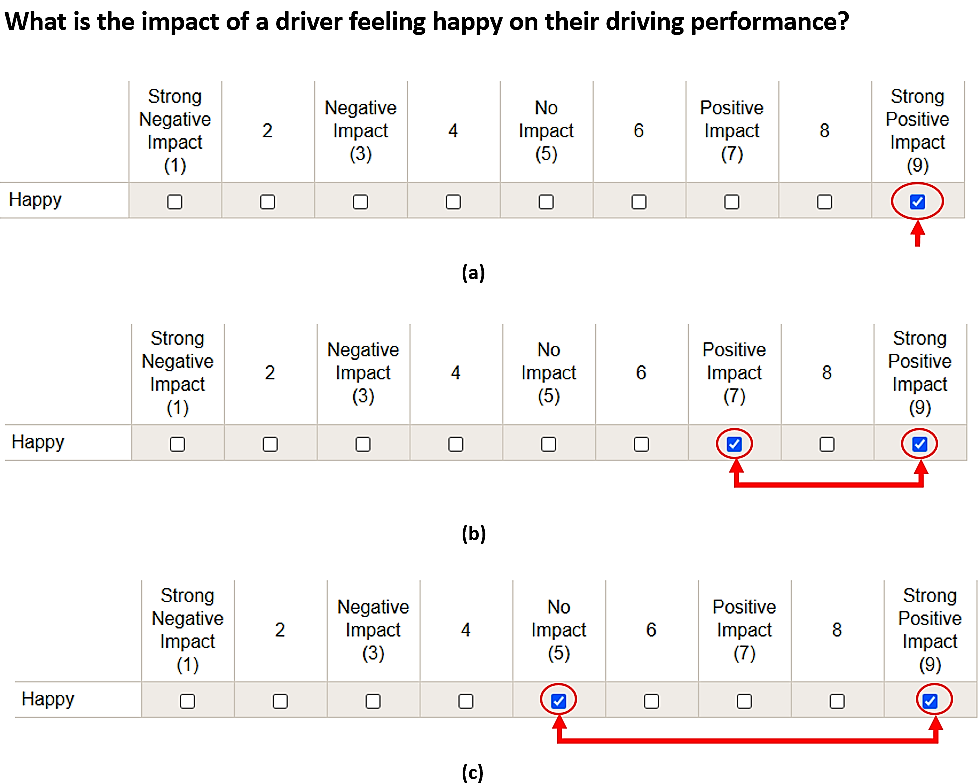}
}
\caption{(a) No uncertainty interval-valued response , (b) Low uncertainty
interval-valued response, and (c) Higher uncertainty interval-valued response.}

\label{fig:uncertexample}
\end{figure}

IAA aims to accurately model information captured through the collection of interval-valued data (e.g., through surveys).
Generally, in IAA, decision makers (experts) are asked to provide their views on factors influencing a certain phenomena by using decision intervals, rather than a single-score answer~~\cite{wagner2014interval,hu2012least}. An interval-valued question consists of participants selecting a range (minimum and maximum values) to represent how certain they are about their responses. 
The width of a given interval denotes experts’ certainty in their answer. For example, for a question assessing the impact of a driver's happiness to their performance (Fig.~\ref{fig:uncertexample}) we could have three scenarios of expert responses. Figure~\ref{fig:uncertexample} (a) shows a response where an expert is certain about the strong  positive impact of a driver feeling happy, therefore giving a single score of $9$. Fig.~\ref{fig:uncertexample} (b) shows a low uncertainty response where an expert is \textit{little} uncertain about the impact of a driver feeling happy, therefore providing an answer with a \textit{narrow} interval of values between $7$ and $9$. Fig.~\ref{fig:uncertexample} (c) represents an overall positive response but with \textit{high} uncertainty interval of the level of impact (between $5$ and $9$).

\begin{table}[b]
\centering
\caption{A sample of collected intervals from three experts (1-3) with the profession \textit{A}}
\begin{tabular}{|c|c|c|c|}
\hline
                               & \textbf{Expert 1} & \textbf{Expert 2} & \textbf{Expert 3} \\ \hline
\textbf{Profession \textit{A}} & {[}1,2{]}         & {[}1,3{]}         & {[}2,4{]}         \\ \hline
\end{tabular}
\label{table:professionA}
\end{table}

IAA utilises collected intervals and generates non-parametric Fuzzy Sets (FSs) to capture all different levels of uncertainty in individual opinions and also between multiple individuals/groups opinions. During IAA FSs model creation, first, the collected intervals are formed in Type-1 FSs (T1 FSs) which minimises the loss of information in experts' opinion. Secondly, these generated T1 FSs are aggregated to z-slice Type-2 General FS (zGT2 FS) which allows to model different individual opinions from different groups of professions all together.

\begin{figure}[h]
\centering
\scalebox{0.32}{
\includegraphics{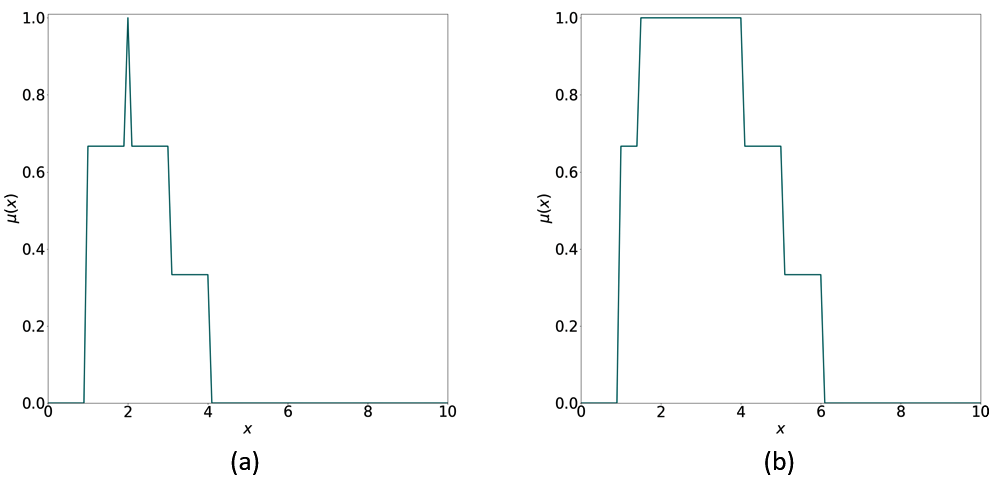}
}
\caption{T1 FSs for each profession using the IAA approach. (a) IAA-Profession A, and (b) IAA-Profession B.}

\label{fig:type1FSs}
\end{figure}

For example, as it is illustrated in Table \ref{table:professionA}, Experts 1-3 -who are working in the same profession \textit{A}- are asked a question and they provide intervals which allow to capture uncertainty in their opinions. In IAA, these collected intervals from each individual are aggregated into a single T1 FS. Thus, the generated T1 FS is able to capture different opinions from experts (Expert 1-3) and model them in a single representation that shows the aggregated opinions of experts from the same profession \textit{A}, as shown in Fig. \ref{fig:type1FSs} (a). The y-axis ($\mu(x)$) represents the level of agreement among the responses e.g. the experts show greatest agreement in their responses at `2'.

Another group of experts (Expert 4-6) -who are working profession \textit{B}- are asked the same questions and they provide different opinions with different levels of uncertainty, as illustrated in Table \ref{table:professionB}. These opinions can be aggregated into another single T1 FSs which is shown in Fig \ref{fig:type1FSs} (b). As can be seen in the comparison of Table \ref{table:professionA} and \ref{table:professionB}, the experts 4-6 tend to be more uncertain about their opinions which leads to a \textit{wider} T1 FS in Fig. \ref{fig:type1FSs}b. 

\begin{table}[!h]
\centering
\caption{A sample of collected intervals from three experts (4-6) with the profession \textit{B}}
\begin{tabular}{|c|c|c|c|}
\hline
                               & \textbf{Expert 4} & \textbf{Expert 5} & \textbf{Expert 6} \\ \hline
\textbf{Profession \textit{B}} & {[}1,5{]}         & {[}1.5,4{]}       & {[}1,6{]}         \\ \hline
\end{tabular}
\label{table:professionB}
\end{table}

\begin{figure}[h]
\centering
\scalebox{0.11}{
\includegraphics{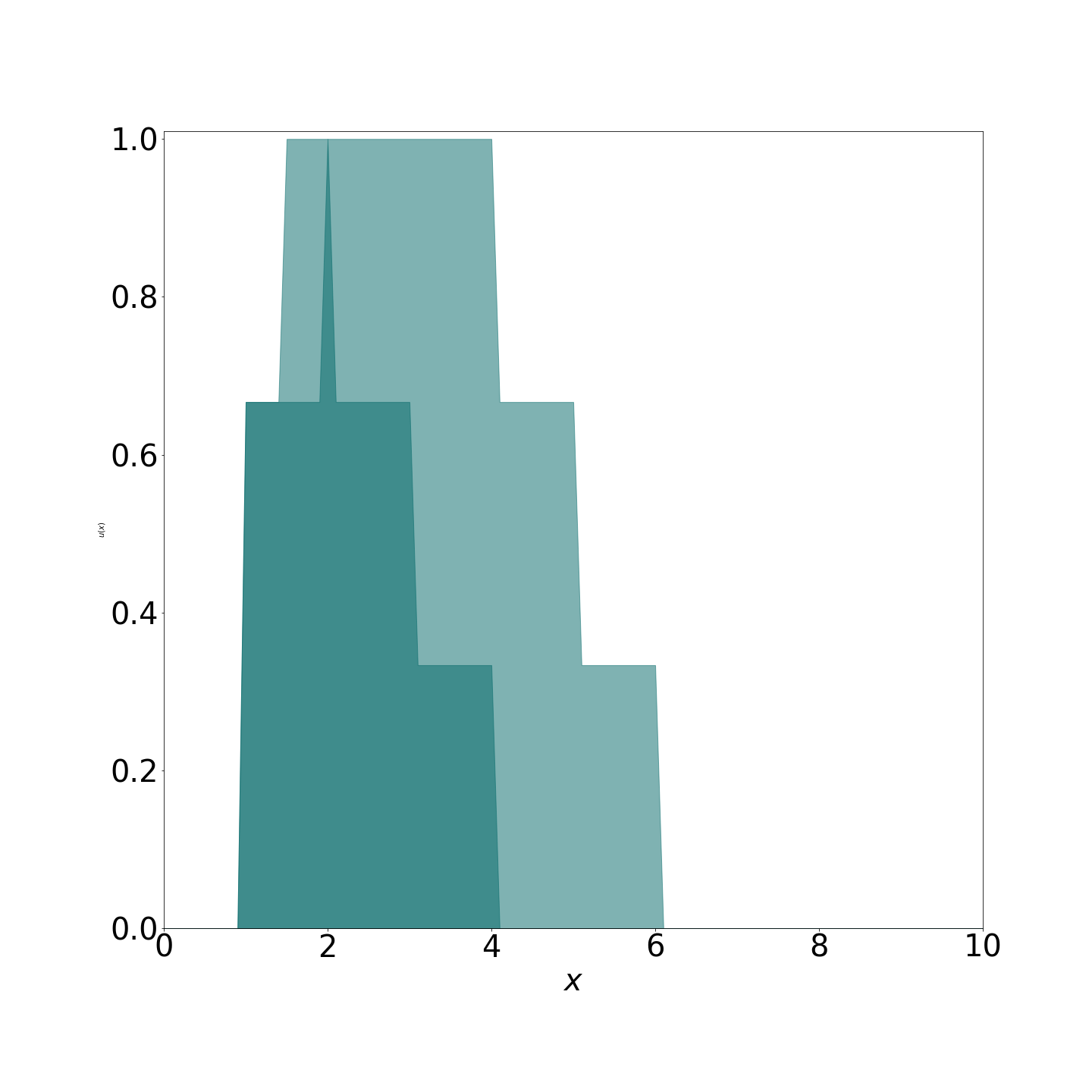}
}
\caption{2D view of the zGT2 FS produced with the IAA T1 FSs from Figure~\ref{fig:type1FSs}}

\label{fig:type2FS}
\end{figure}

Later, the created T1 FSs are aggregated to generate zGT2 FSs where the agreement/variation between multiple experts/groups of information is modelled through the secondary memberships (zslices), as demonstrated in Fig. \ref{fig:type2FS}. The darker area in the plot represents the region with higher agreement between the two IAA FSs i.e. both groups agree the most in their responses between 1 and 4. 

For a detailed description of zGT2 FSs construction, please see the respective study in~\cite{wagner2014interval}. The python libraries and functions used in this study to develop the FSs can be found in~\cite{mcculloch2017fuzzycreator}.


\begin{figure*}[!htpb]
\centering
\scalebox{0.46}{
\includegraphics{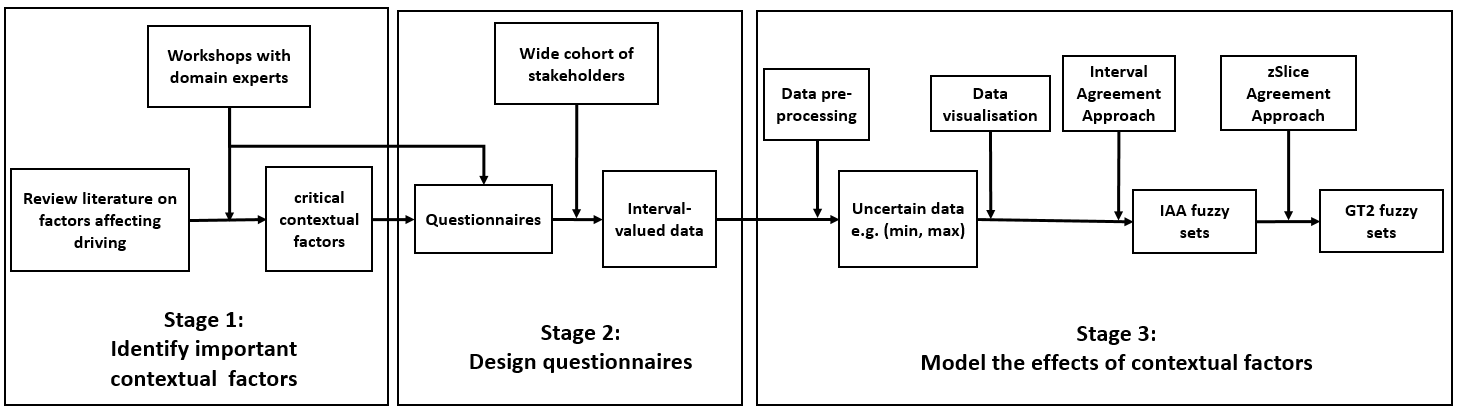}
}
\caption{Our novel methodology to capture and understand the effects of contextual factors on commercial driving using a wide cohort of experts.}

\label{fig:method}
\end{figure*}

\section{Methodology}
\label{method}
This section introduces our novel approach, EDA. EDA identifies, captures, analysis and models the effects of contextual factors on driving performance to be embedded into intelligent decision making systems. The methodology consists of the following stages:

\subsection{Stage 1: Identify important contextual factors}
\label{identifying}

This stage involves compiling a list of contextual factors that affect driving performance based on literature. Subsequently, we organise consultative workshops with domain experts who have experience in the industry and road safety regulations. During the workshops, we present the factors obtained from the literature and ask the experts to share their opinions on two questions: ``Are the contextual factors identified from the literature sensible and valid in the current driving environment?" and ``Is there any other contextual factor we should consider?" These workshops are conducted to validate, update and prioritise the factors obtained from the literature as some of the factors may already be outdated due to advances in technologies and road safety regulations. In addition, the experts assist in identifying the different types of stakeholders in the industry (i.e. drivers, managers, researchers, and road safety professionals), who will provide insights about the effects of the factors. 

\subsection{Stage 2: Designing questionnaires}
\label{design}

After identifying and validating the factors in Stage 1, we proceed to design interval-valued response-format questions (as described in Section~\ref{iaa}) to collect the opinions of experts from the different types of stakeholders identified in Stage 1 (also know as professions in this paper). We propose interval-valued questions as they are powerful and natural means for participants to communicate indecision or imprecision in their opinions~\cite{ellerby2019decsys}. The design and wordings of the questions are improved during the workshops with the domain experts recruited in Stage 1. The experts are asked the following questions: ``Is the format and rating scale of the questionnaire easy to comprehend and complete?", ``Are the wordings and questions understandable?", ``Is the length of the questionnaire adequate considering the busy schedules of professionals?", and ``Is there anything important missing in the questionnaire?". The final questionnaire is distributed to a wide cohort of stakeholders to obtain their views about the impact of the contextual factors.

\subsection{Stage 3: Understanding the factors and modelling their impact}
In this stage, we gather the responses from the questionnaire and transform each response into (minimum, maximum) interval format representing the range of a response. For example, the responses for the questions in Fig.~\ref{fig:uncertexample} will be transformed into tuples to represent the effect of `happy' i.e.,  \textit{(9,9)} for the response in (a), \textit{(7,9)} for (b), and \textit{(5,9)} for (c). This transformation is important for efficient visualisation and development of fuzzy sets. We adopt the least commitment strategy~\cite{hu2012least} by using all responses in our analysis (i.e. no outlier removal) as outliers may contain rich information and IAA fuzzy sets are able to efficiently handle this information.

The transformed data are visualised using line graphs and box-plots to understand the individual as well as group opinions respectively. We are interested in understanding: 1) the difference in opinions across the different professions, 2) the level of certainty in the responses of experts, 3) agreement amongst experts within each profession, and 4) most importantly, the final impact of factors by combining the opinions of all experts. 

We establish IAA fuzzy sets from the interval-valued responses of each factor in each profession. The fuzzy sets account for any variability in the responses of experts. Therefore, if there exist $n$ contextual factors and $i$ professions, $n \times i$ IAA fuzzy sets will be generated. We then use fuzzy set similarity measure e.g., Jaccard similarity measure~\cite{wagner2013similarity}, to quantitatively express the level of agreement among the opinions of the different stakeholders. 

To obtain the final impact of the factors on driving performance, we aggregate the IAA FSs into zGT2 FSs~\cite{wagner2014interval} by employing the agreement principle in Wagner \textit{et al}~\cite{wagner2011employing} and associating a higher secondary membership (zLevel) to areas where the IAA FSs overlap. That is, if $n \times i$ IAA FSs are generated for $n$ contextual factors and $i$ professions, each FS representing a specific factor is aggregated with their corresponding FSs to produce $n$ zGT2 FSs. The secondary membership captures the agreement among the different professions and the zGT2 FSs provide a clear representation and separation of the individual types of uncertainty present in the data (as described in Section~\ref{uncertknowledge}). The resulting zGT2 FSs will be integrated into intelligent driver-assessment systems (as shown in Section~\ref{application eda} below).

\section{Case Study}
\label{experiment}
In this study, we apply EDA to HGV driving in the United Kingdom (UK) due to the importance of HGVs in delivering goods and services across the nation. First, we identify and capture the effects of contextual factors from HGV professionals. Later, the collected experts' opinions are modelled and embedded into an intelligent HGV driver-assessment system to automatically moderate decisions and outcomes. We present the experimental design of our study in the following sections. 

\subsection{Workshops with HGV domain experts}
First, we identified the important contextual factors of HGV driving from the literature. Next, we organised five iterative workshops with domain experts to validate the contextual factors identified from the literature and the questionnaire design. In each iteration, the domain experts refined their responses to the questions presented in Sections~\ref{identifying} and~\ref{design} about the influencing factors, questionnaire design, format and instructions. These workshops were held virtually due to COVID-19 restrictions between September 2020 and November 2020. We recruited nine domain experts consisting of a university professor in Psychology specialised in driving behaviour, three HGV fleet managers, and five researchers specialised in intelligent driver assessment systems. We believe nine participants are sufficient because too many participants would not be manageable, as there would be too many opinions and potentially too much noise. To complement the lack of HGV drivers in the workshops, we interviewed the first participants who completed the questionnaire i.e. two HGV drivers, two researchers and two road safety professionals, asking them if there were any other important factors missing from the questionnaire and whether the design of the questionnaire was appropriate. Using their responses, we updated the questionnaire.

\subsection{Questionnaire}
\begin{figure}[!htpb]
\centering
\scalebox{0.45}{
\includegraphics{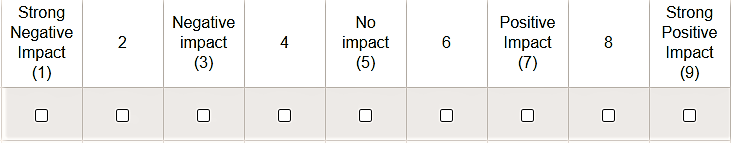}
}
\caption{Rating scale for questions in our study}

\label{fig:ratingscale}
\end{figure}

Our questionnaire consisted of interval-valued questions that asked participants to provide their expert opinions or ratings about the impact of the critical factors (elicited in the previous stages) on HGV  driving performance. Clear guidelines and instructions on how to answer the questions and the purpose of using a nine-point interval-valued scale were provided i.e. the possibility for participants to select two points representing the range of certainty of their responses. The nine-point rating scale ranged from 1, meaning `strong negative impact', to 9, `strong positive impact' and 5 representing `no impact' as shown in Fig.~\ref{fig:ratingscale}.

\subsection{Participant recruitment}
\begin{figure}[!t]
\centering
\scalebox{0.25}{
\includegraphics{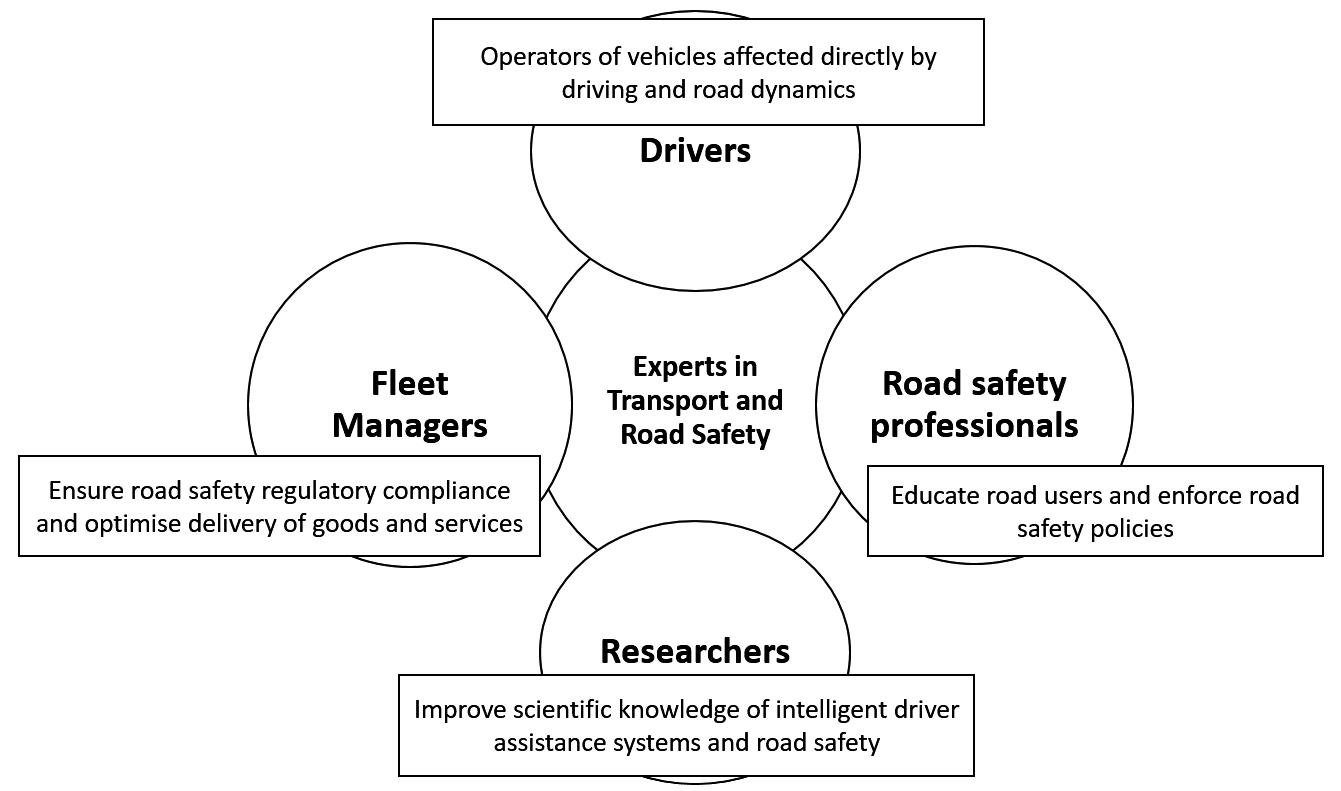}
}
\caption{Diagram showing the different types of stakeholders in the HGV industry along with the definitions of their roles.}
\label{fig:experts}
\end{figure}

The four stakeholders identified during the workshops with the domain experts were fleet managers, road safety professionals, HGV drivers and researchers. Fig~\ref{fig:experts} defines the roles of each stakeholder. HGV drivers are operators of the vehicles and the main beneficiary of this study as we intend to understand the influence of contextual factors to moderate the evaluation of their driving behaviours. Road safety professionals enforce road safety regulations on road users and the results of this study will assist them in developing adequate traffic laws that take into consideration the contextual factors.  Researchers improve scientific knowledge of intelligent driver assistance systems and road safety. The findings of this study will assist them in prioritising the detection of high impact factors and optimising the regulation of decisions. Fleet managers ensure their companies are compliant with road safety regulations and manage drivers to optimise delivery of goods and services. These managers will benefit from this study by adapting their driver-assessment systems to include context. In addition, the managers can organise workshops with their drivers regarding factors in which the managers and drivers disagree in opinions. 

Participants for our study were recruited by sending mass messages to individuals on LinkedIn~\cite{linkedin}, University of Nottingham and Transport Study Group (UTSG) whose job titles and expertise matched any of the different stakeholders. No compensation was offered for participation.

\subsection{Sample}
Table~\ref{tab:part} provides a summary of the number of participants in the study and the average years of experience. Ninety-three participants from the UK completed the questionnaire. Among the participants were: 20 HGV drivers, 23 researchers, 24 fleet managers, and 26 road safety professionals. Years of experience ranged from 3 to 46 years with average and standard deviations of (M=22.90, SD=12.79) for HGV drivers, (M=17.33, SD=10.72) for researchers, (M=11.04, SD=8.22) for fleet managers, and (M=20.81, SD=12.39) for road safety professionals. 

\begin{table}[htbp]
  \centering
  \caption{Distributions of participants, average years of experience, and standard deviations of experience among the groups of stakeholders}
  \scalebox{0.73}{
    \begin{tabular}{|l|c|c|c|c|}
    \hline
          & \textbf{HGV drivers} & \textbf{Researchers} & \textbf{Fleet managers} & \textbf{Road Safety Prof} \\
  \hline
    \textbf{No. of Participants} & 20    & 23    & 24    & 26 \\
\hline
    \textbf{Avg. Experience (Yrs)} & 22.90 & 17.33 & 11.04 & 20.81 \\
\hline
    \textbf{Std. Experience (Yrs)} & 12.79 & 10.72 & 8.22  & 12.39 \\
\hline
    \end{tabular}%
    }
  \label{tab:part}%
\end{table}%

\subsection{Model selection and settings}
We use IAA and zGT2 FSs~\cite{wagner2014interval} to model the responses and variability in opinions among experts in the different professions. We employ Jaccard similarity measure~\cite{wagner2013similarity} to calculate the agreement in opinions amongst the different professions. The Jaccard similarity measure is an efficient and well-established method used to calculate similarity between fuzzy sets. It calculates the cardinality of the intersection of two sets, divided by the cardinality of the union of the two sets. The output value for the method lies between 0 and 1, where 1 indicates total agreement and 0 indicates total disagreement. 



\section{Case Study Results}
\label{results}
\subsection{Identification of contextual factors}
\begin{figure}[!htpb]
\centering
\scalebox{0.56}{
\includegraphics{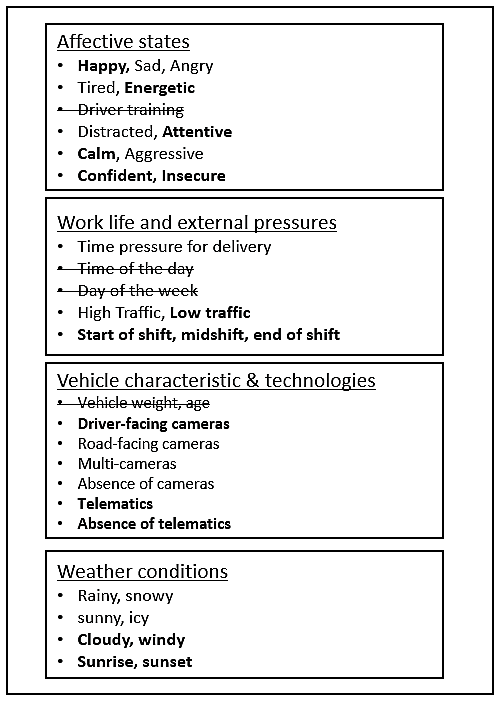}
}
\caption{Contextual factors that affect HGV driving performance extracted from the literature and validated by experts. The strikethrough factors were eliminated by experts during a workshop while the bold texts represent additional factors identified by the experts.}

\label{fig:litreview}
\end{figure}

Contextual factors identified from the literature~\cite{mahajan2019effects,foy2018mental,stern2019data,wu2018influence,kidd2016influence,mase2020evaluating,agrawal2019towards} (see Fig.~\ref{fig:litreview}) were presented to the experts in the workshops for validation. Some factors were identified as irrelevant, such as vehicle characteristics, while others were identified as outdated due to new road safety policies in the UK, such as rest breaks. Other factors, such as time of the day and day of the week were revised to start, mid and end of shifts, as HGV drivers start their jobs at different times of the day and different days of the week. The experts proposed additional positive related factors, driver affective states and weather conditions that could affect drivers. Fig.~\ref{fig:litreview} shows the updated list of critical factors included in our study. Those eliminated during the workshop are represented using strikethrough texts; additional factors arising from our discussions with experts are in bold text. 17 additional factors to those found in the literature were identified as a result of EDA stage 1 (Section~\ref{identifying}). A total of 31 factors were utilised to design the interval-valued questionnaire\footnote{Survey link: https://nottingham.onlinesurveys.ac.uk/driving-performance}.

\subsection{Effects of contextual factors on HGV drivers' performance}
We used box-plots and line graphs to visualise the responses from experts. The box-plots show the group distributions of responses while the line graphs show the individual responses. Each line or dot in the line graphs represents the \textit{(minimum, maximum)} response of each expert and the colours represent the different professions. If an expert's response was very certain (i.e., they selected a single value as their rating), a single point is plotted in the line graph. Each line graph has 93 lines and points for all 93 experts. The line graphs are found in the supplementary results section (Appendix). We show the mode rather than the median in the box-plot distributions as we are interested in the rating provided by majority of the experts in each profession. We use the acronyms HD=HGV Driver, FM=Fleet Manager, R=Researcher, RS= Road Safety professional, to represent the different professions on the box-plots. 

\begin{figure*}[!htpb]
\centering
\scalebox{0.7}{
\includegraphics{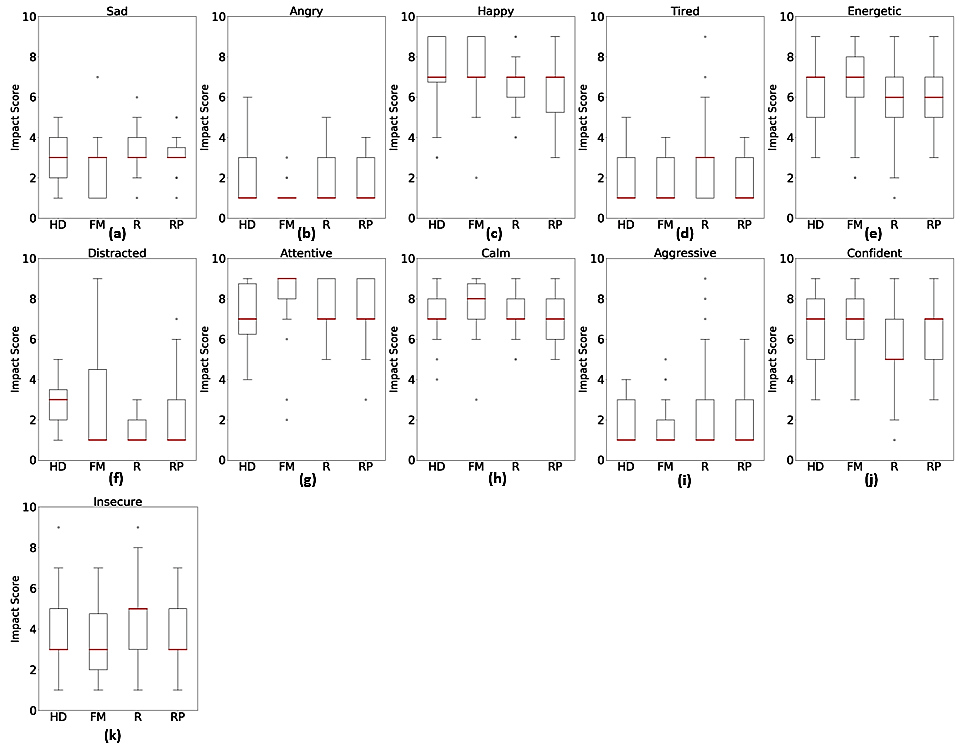}
}
\caption{Box-plots showing the distribution of responses from drivers, fleet managers, researchers and road safety professionals about the impact of drivers' affective states on their driving.}

\label{fig:affect_boxplot}
\end{figure*}

\subsubsection{\textbf{Driver affective states}}
The box-plots in Fig.~\ref{fig:affect_boxplot},  reveal a less negative effect of feeling tired (Fig.\ref{fig:affect_boxplot}d) suggested by researchers. This may be due to new road safety regulations in the UK that require drivers to take frequent rest breaks~\cite{meng2015driving}. Next, we notice that the majority of researchers suggest that being confident (Fig.\ref{fig:affect_boxplot}j) or insecure (Fig.\ref{fig:affect_boxplot}k) has no impact on drivers' performance (i.e. mode = 5), which contradicts what majority of experts from the other professions think. The other experts suggest that being confident has a positive impact, while being insecure has a negative impact on performance. 

The box-plots show that experts agree on how affective states influence driving (i.e. positively or negatively). However, we observe variations among professions with regards to the intensity of the influence of those factors. These variations can be attributed to the distinct roles, knowledge, experiences and goals of stakeholders. Such variations in opinions need to be considered in the development of intelligent driver-assessment systems to ensure that the systems are not biased towards particular stakeholders. This is the main objective of stages 2 and 3 of EDA, where we collect responses from a wide cohort of stakeholders and incorporate the consensus among their views into intelligent systems. 

Although not always easy to capture, some of the affective states brought out by experts during our workshops can be automatically detected using machine learning. Deep learning approaches applied to driver-facing footage, for instance, have shown promising results in automatically identifying some of the affective states, such as, distracted or attentive driving~\cite{mase2020benchmarking,eraqi2019driver}, different types of human emotions~\cite{jain2018hybrid,sanchez2021affective}, and tired or energetic~\cite{ed2019driver,cui2021real}. Alternatively, calm or aggressive driving is accurately detected using telematics incident data~\cite{agrawal2019towards,mase2020capturing}, while more complex affective states such as confidence and insecurity are still  difficult to detect. 

\subsubsection{\textbf{Work-life factors}}
\begin{figure*}[h]
\centering
\scalebox{0.4}{
\includegraphics{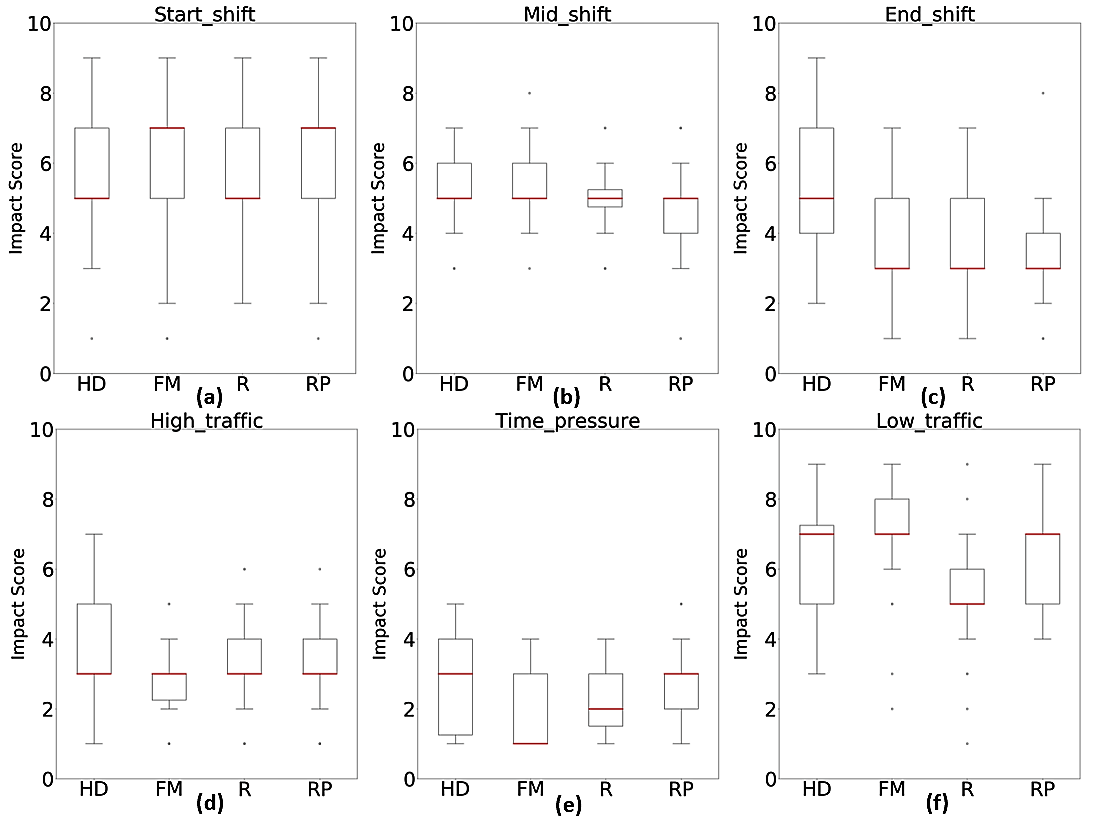}
}
\caption{Box-plots showing the distribution of responses from drivers, fleet managers, researchers and road safety professionals about the impact of work related factors.}

\label{fig:work_boxplot}
\end{figure*}

Fig.~\ref{fig:work_boxplot} depicts the distributions of responses from experts about the impact of work-life factors. We observe variation between the opinions of drivers compared to fleet managers and road safety professionals for start and end of a shift (Fig.\ref{fig:work_boxplot}a,c) with the mode of drivers at 5. The majority of fleet managers and road safety professionals indicate that the start of a shift has a positive influence on driving and the end of a shift has a negative influence. To better understand the cause of this variability, further investigation and interviews with the different stakeholders is required. Furthermore, time pressure for delivery (Fig.\ref{fig:work_boxplot}e) is considered to have a strong negative impact (mode = 1) by the majority of fleet managers, who sometimes exert pressure on drivers to deliver on time~\cite{mic2020}. This observation stresses the need for moderation of driver's performance, as HGV drivers face pressure from their companies to deliver goods on time, leading to accidents and deaths~\cite{mcghee2020,mic2020}. It is unfair to ignore the pressure from their employers when assessing  driving behaviours, as their jobs are on the line if they do not comply.

The time of shift (i.e. start, mid and end of shift) can be automatically identified using job dispatch and routing management systems, while traffic state (i.e. high or low) can be automatically detected from road-facing camera images using computer vision techniques~\cite{chakraborty2018traffic,agarwal2021camera} or obtained from location based systems e.g., Google Maps.

\subsubsection{\textbf{In-vehicle technologies}}
\begin{figure*}[h]
\centering
\scalebox{0.4}{
\includegraphics{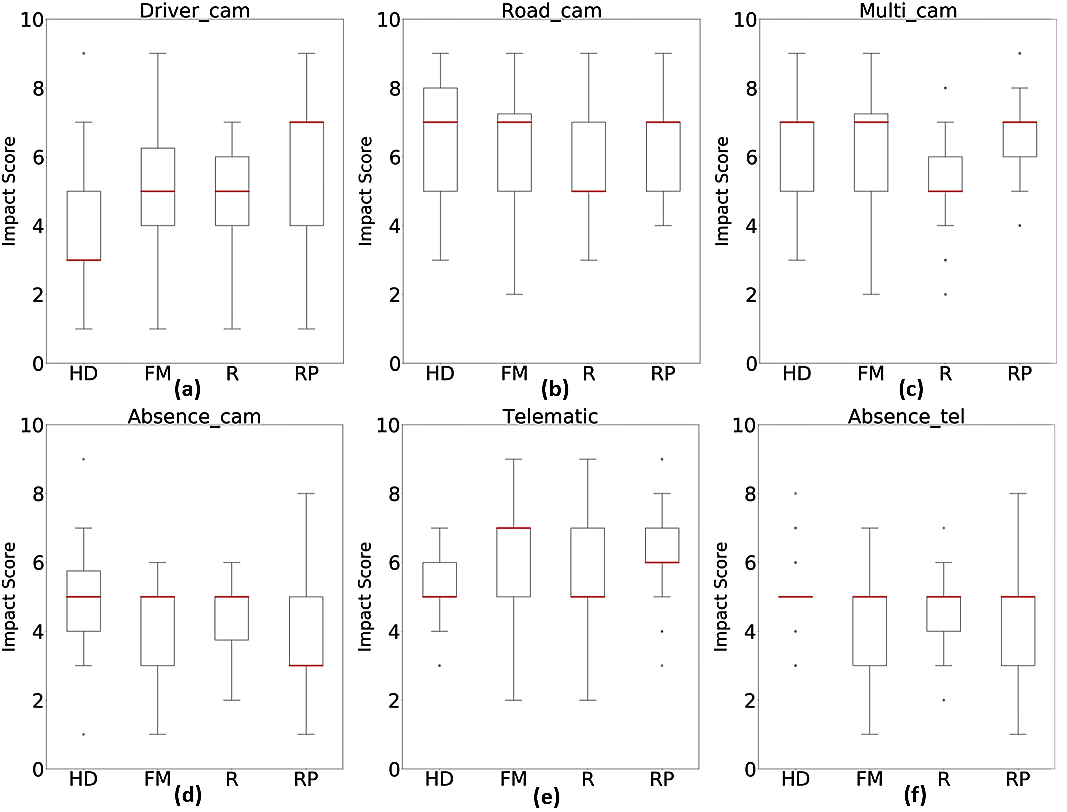}
}
\caption{Box-plots showing the distribution of responses from drivers, fleet managers, researchers and road safety professionals about the impact of technologies.}

\label{fig:tech_boxplot}
\end{figure*}

Fig.~\ref{fig:tech_boxplot} depicts the distributions of responses from experts about the impact of in-vehicle technologies. Drivers and road safety professionals disagree about the impact of driver-facing cameras (Fig.\ref{fig:tech_boxplot}a) as majority of drivers suggest driver-facing cameras have a negative impact on driving while road safety professionals believe driver-facing cameras have a positive impact. In addition, road safety professionals suggest that the absence of cameras (Fig.\ref{fig:tech_boxplot}d) has a negative impact on driving while majority of other stakeholders suggest no impact. The negative impact of driver-facing cameras suggested by drivers may be a consequence of how the videos or images are used (e.g. used to penalise drivers) or due to privacy concerns as we clearly observe positive ratings by drivers for road-facing cameras, which are less intrusive and personal. 

A moderation system developed with only the opinions of drivers may be inaccurate with regards to the effects of driver-facing cameras, as work such as Mase \textit{et al}~\cite{mase2020evaluating} shows a significant reduction in driving incidents due to both driver-facing and road-facing cameras. Such discrepancies in opinions can only be identified when different stakeholders are considered; the motivation of stage 2 of EDA. The discrepancies can be resolved by organising follow-up interviews with stakeholders to understand their viewpoints or by combining all opinions and finding a consensus (i.e. stage 3 of EDA).

\begin{figure*}[h]
\centering
\scalebox{0.4}{
\includegraphics{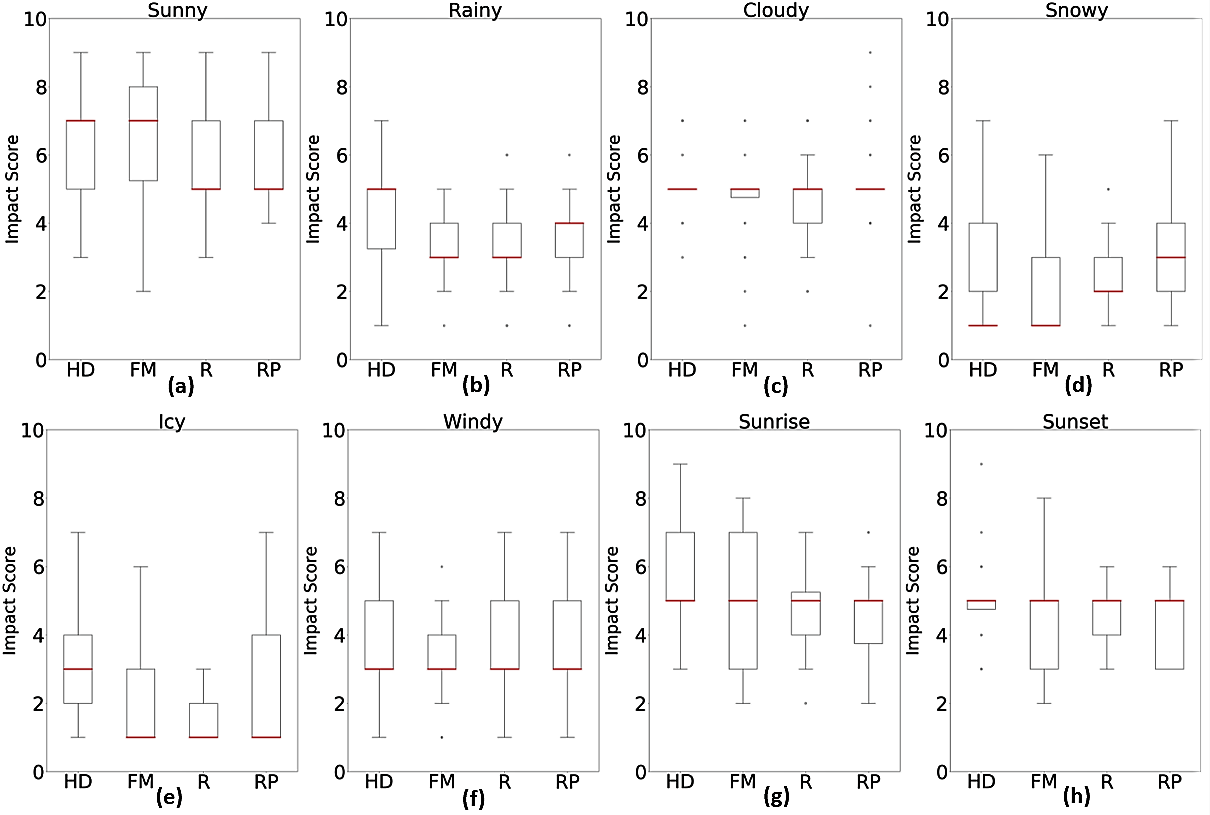}
}
\caption{Box-plots showing the distribution of responses from drivers, fleet managers, researchers and road safety professionals about the impact of weather conditions.}

\label{fig:weather_boxplot}
\end{figure*}


\subsubsection{\textbf{Weather conditions}}
Fig.~\ref{fig:weather_boxplot} presents the distributions of responses on the impact of weather conditions. The majority of drivers and fleet managers suggest a sunny weather (Fig.\ref{fig:weather_boxplot}a) has a positive impact (mode = 7) while majority of researchers and road safety professional suggest it has no impact(mode = 5). Similarly, the majority of drivers indicate a rainy weather (Fig.\ref{fig:weather_boxplot}b) has no impact on driving, while the other stakeholders believe it has a negative impact. The weather conditions identified in this study can be automatically detected using deep learning methods on road-facing images~\cite{ibrahim2019weathernet,xia2020resnet15,zhao2018cnn} or obtained from online weather data sources such as the Metropolitan Police UK\footnote{https://www.met.police.uk/}.

\subsection{Agreement amongst the different stakeholders}
\begin{figure*}[h]
\centering
\scalebox{0.38}{
\includegraphics{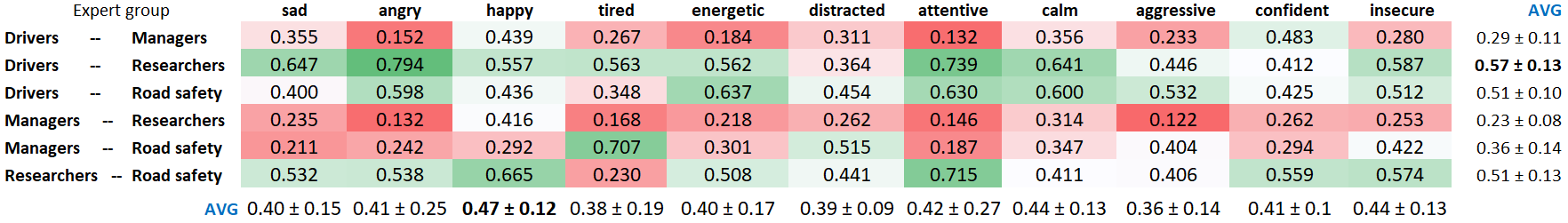}
}
\caption{Agreement amongst stakeholders opinions for affective factors}

\label{fig:simaffect}
\end{figure*}

\begin{figure*}[h]
\centering
\scalebox{0.4}{
\includegraphics{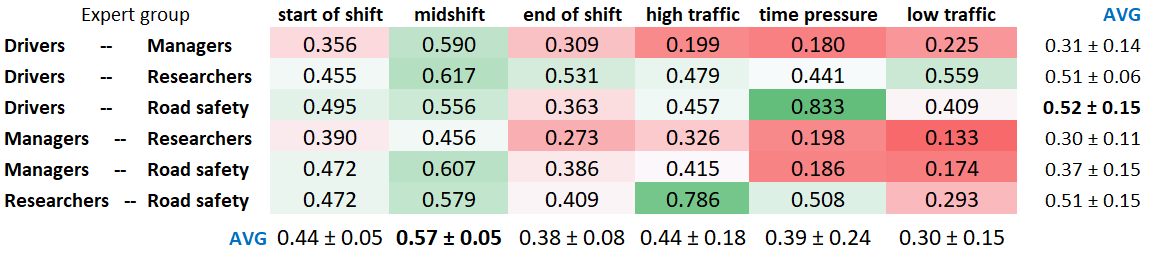}
}
\caption{Agreement amongst stakeholders opinions for work-life factors}

\label{fig:simwork}
\end{figure*}

\begin{figure*}[h]
\centering
\scalebox{0.4}{
\includegraphics{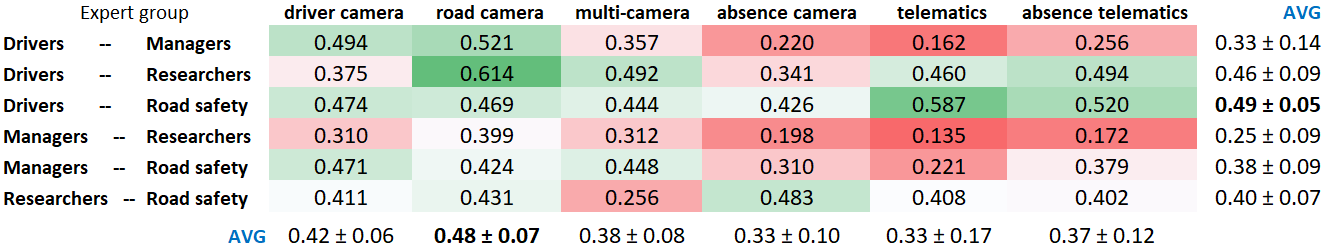}
}
\caption{Agreement amongst stakeholders opinions for in-vehicle technologies}

\label{fig:simtech}
\end{figure*}

\begin{figure*}[h]
\centering
\scalebox{0.4}{
\includegraphics{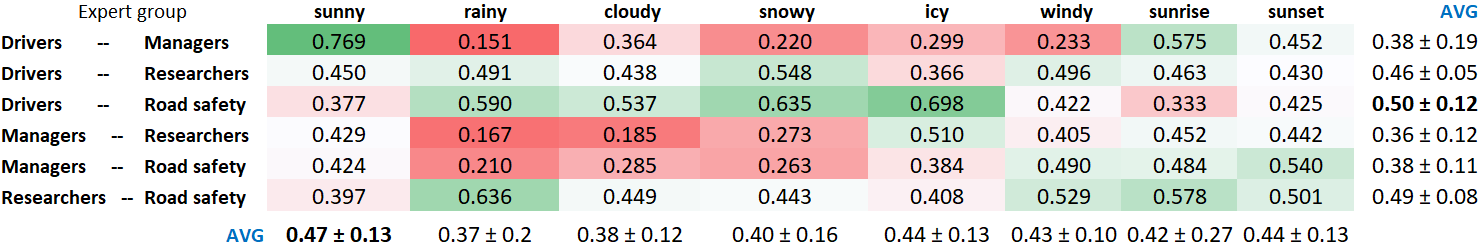}
}
\caption{Agreement amongst stakeholders opinions for weather conditions}

\label{fig:simweath}
\end{figure*}

In order to validate our rationale of using different types of stakeholders to provide the effects of contextual factors (stage 2 of EDA), we calculate the similarity measures between the different professions for each contextual factor to show the agreement among opinions. The results are presented in Fig.~\ref{fig:simaffect},~\ref{fig:simwork},~\ref{fig:simtech} and~\ref{fig:simweath} with similarity values between 0 and 1, where 0 indicates total disagreement (red) and 1 indicates total agreement (green). We observe from the figures that drivers, researchers and road safety professionals have similarity values above 0.5 for several factors.  However, we observe low levels of agreement (below 0.5) between drivers and researchers in 18 out of 31 factors, with the highest disagreement in the absence of camera, and 17 out of 31 factors between drivers and road safety professionals. Similarly, we identify 17 out of 31 factors between researchers and road safety professionals with agreement values less than 0.5. Furthermore, fleet managers disagree the most with other stakeholders in the following number of factors: 27 between drivers, 30 between researchers, and 27 between road safety professionals. 

\begin{figure*}[!htpb]
\centering
\scalebox{0.6}{
\includegraphics{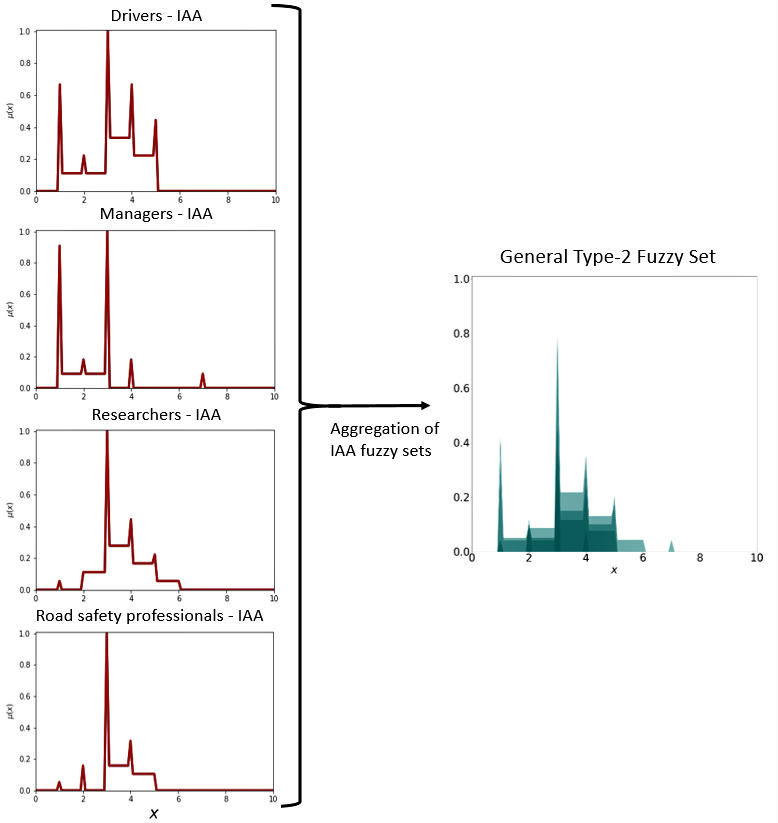}
}
\caption{The aggregation of IAA fuzzy sets for the impact of drivers feeling `sad' on HGV driving obtained from drivers, fleet managers, researchers and road safety professionals}

\label{fig:interiaa}
\end{figure*}

The perceived large number of disagreements among stakeholders, especially between fleet managers and other stakeholders, validates our hypothesis of considering different stakeholders. However, by considering several stakeholders to provide the influence of contextual factors, we are faced with the question, `how do we combine the effects of factors with high disagreement among stakeholders?'. In order to answer this question, we first develop IAA FSs for the responses of experts from each profession. For instance, Fig.~\ref{fig:interiaa} shows IAA FSs of `feeling sad' for HGV drivers, fleet managers, researchers and road safety professionals. The IAA FSs model and provide a clear visualisation of the variability in opinions among experts within the same profession. Examining the IAA FSs of `feeling sad' in Fig.~\ref{fig:interiaa}, we observe that opinions of experts in all four professions have some level of uncertainty about the impact of feeling sad on HGV driving e.g. drivers' IAA FS ranges from 1 to 5 with the highest agreement at 3 (i.e. maximum of the FS), while managers' IAA FS ranges from 1 to 4 with an outlier at 7. Also, the FS of managers is more skewed compared to the FS of drivers indicating that more managers suggest `feeling sad' has a strong negative impact compared to drivers.  

\begin{figure*}[!htpb]
\centering
\scalebox{0.48}{
\includegraphics{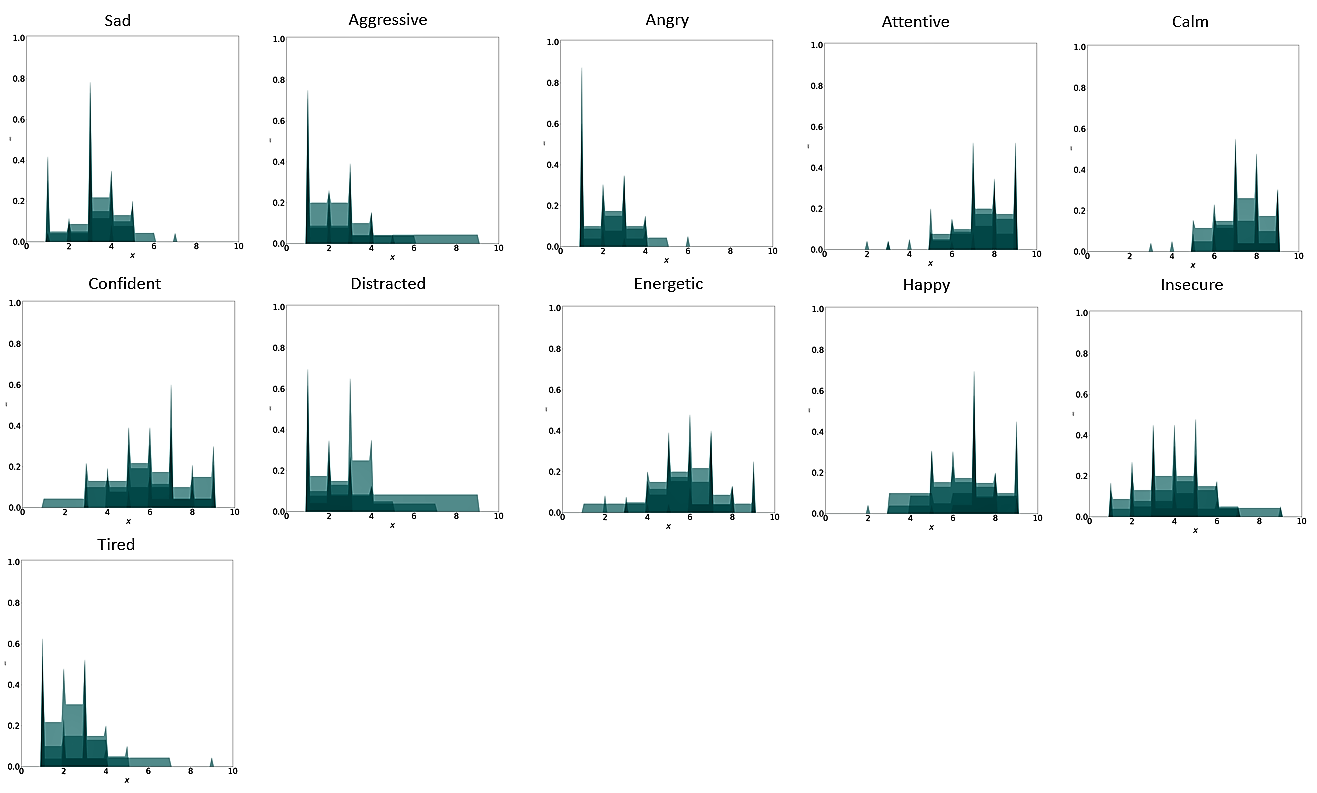}
}
\caption{Resulting General Type-2 Fuzzy Sets of affective states showing variability in viewpoints among experts}

\label{fig:effect_affect}
\end{figure*}

\begin{figure*}[!htpb]
\centering
\scalebox{0.46}{
\includegraphics{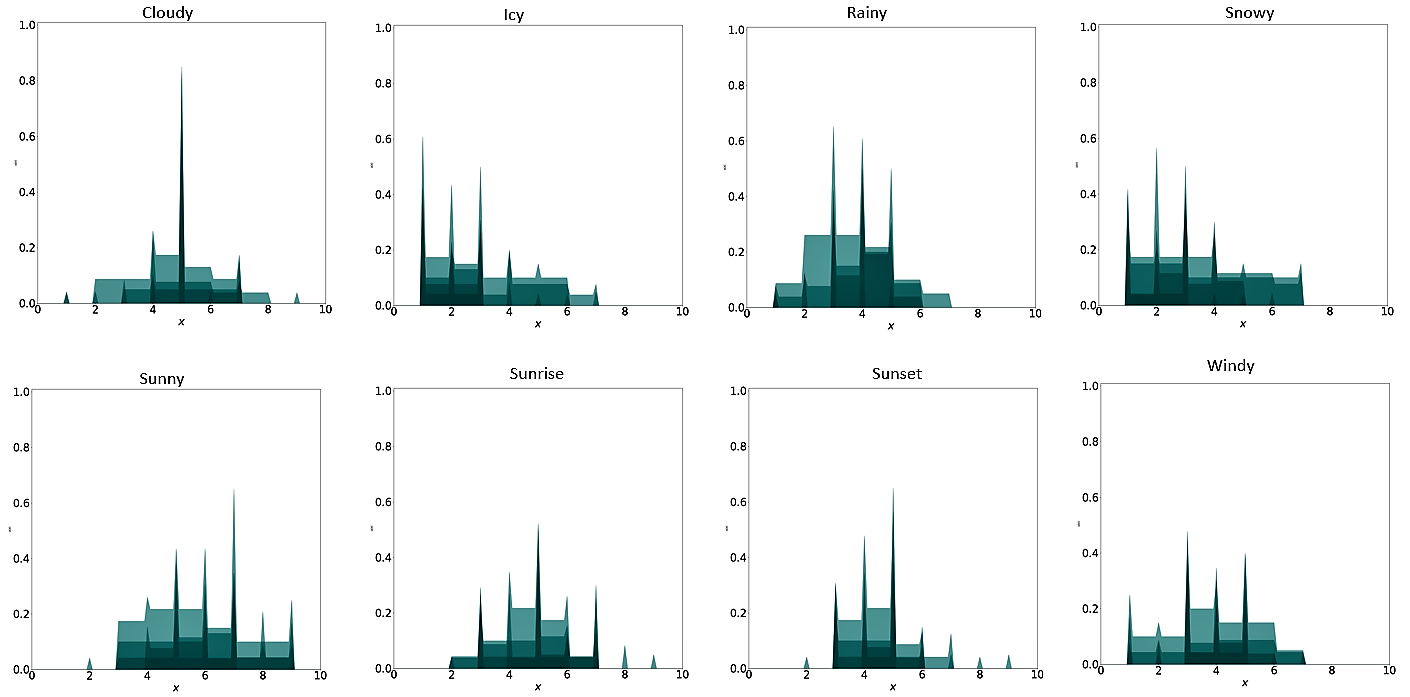}
}
\caption{Resulting General Type-2 Fuzzy Sets of weather conditions showing variability in viewpoints among experts}

\label{fig:effect_weather}
\end{figure*}

\begin{figure*}[!htpb]
\centering
\scalebox{0.5}{
\includegraphics{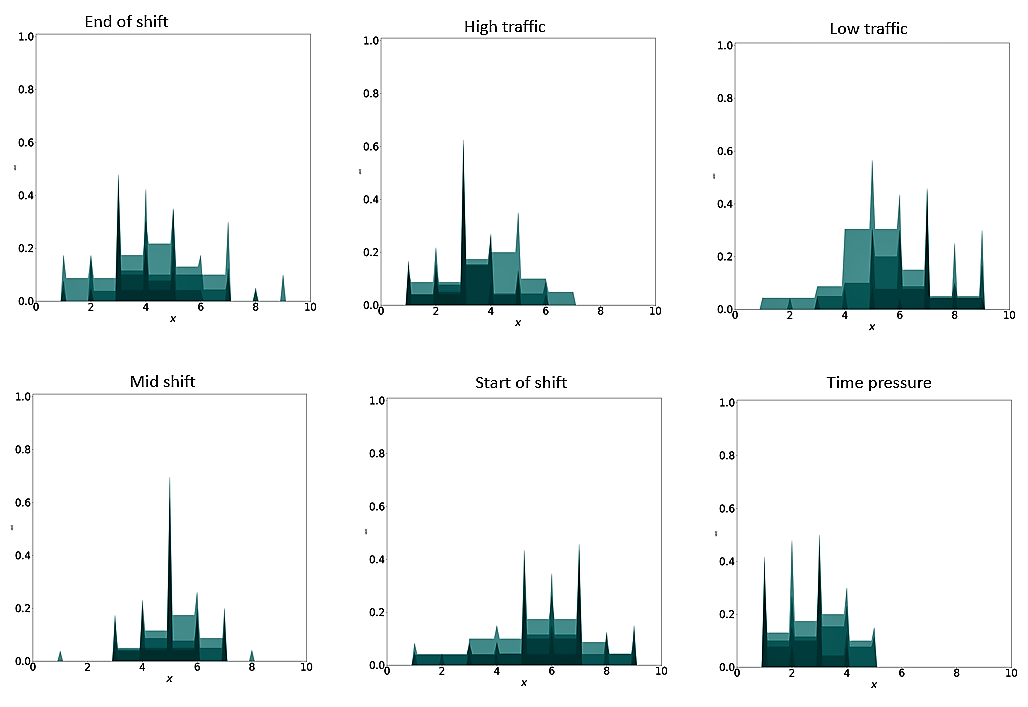}
}
\caption{Resulting General Type-2 Fuzzy Sets of work-life factors showing variability in viewpoints among experts}

\label{fig:effect_work}
\end{figure*}

\begin{figure*}[!htpb]
\centering
\scalebox{0.48}{
\includegraphics{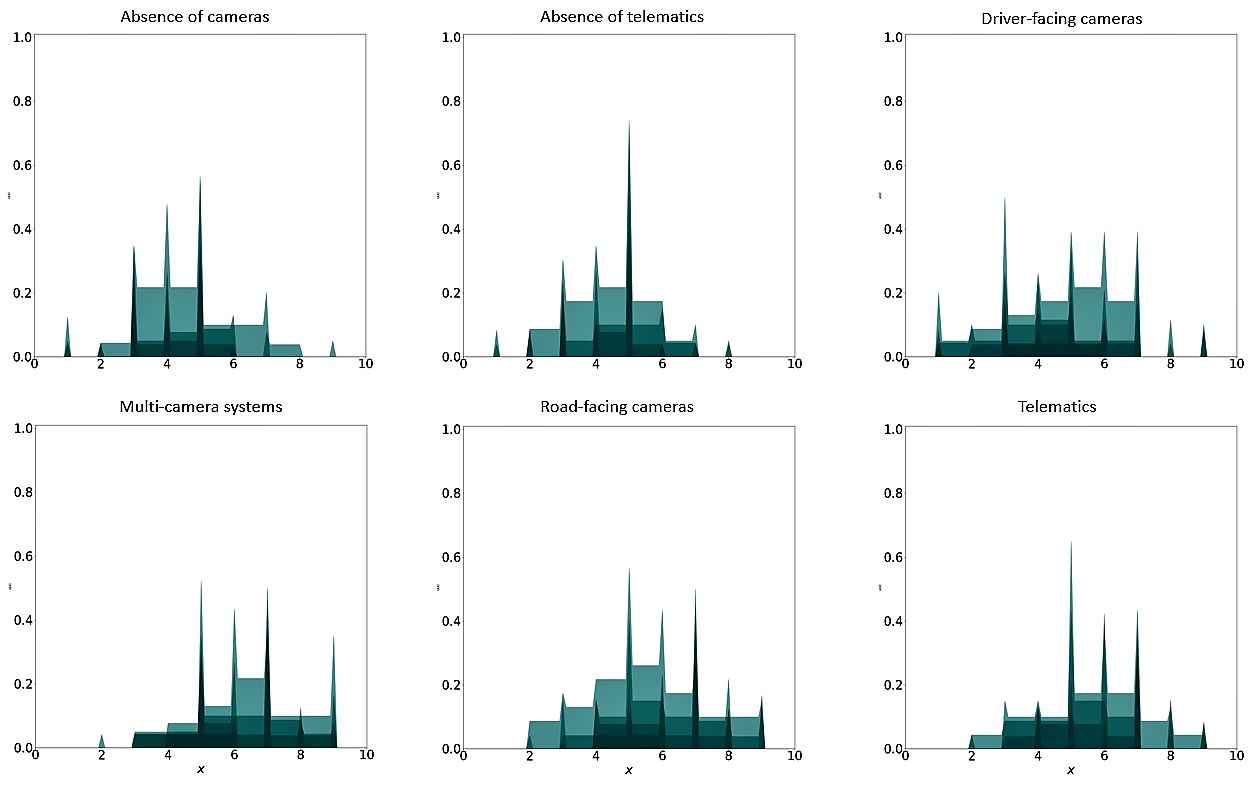}
}
\caption{Resulting General Type-2 Fuzzy Sets of in-vehicle technologies showing variability in viewpoints among experts}

\label{fig:effect_tech}
\end{figure*}

To obtain the final FSs representing the overall impact of factors, we aggregate the IAA FSs into zGT2 FSs as shown in Fig.~\ref{fig:interiaa}. By adopting the least commitment strategy of information fusion~\cite{hu2012least}, we assume equal weights for all stakeholders and experts to enable easy adaptation of our framework in the future. The resulting zGT2 FSs (Fig.~\ref{fig:effect_affect}, \ref{fig:effect_weather}, \ref{fig:effect_work} and \ref{fig:effect_tech}) represent the agreement and variability of all experts' opinions across all professions. The advantage of using the IAA approach in EDA is : 1) its generalisability to any interval-valued dataset as no information is added or removed, and assumptions are kept to a minimum such as the membership function type e.g. Gaussian or Triangular, and 2) it provides interpretable representations of expert knowledge and uncertainty levels. The shaded regions in the zGT2 FSs represent the areas where the IAA FSs overlap, effectively weighting areas with high agreement among professions. Dark areas represent high agreement among professions. Thus, as shown in Fig.~\ref{fig:interiaa}, the outlier at `7' is given a low weight denoted by the bright shade. This approach of weighting the impact of factors according to the agreement among stakeholders illustrated by the zGT2 FSs provides a logical and visual solution to combine the opinions of different stakeholders even with high disagreements. 

\subsection{Embedding EDA into intelligent systems}
\label{application eda}

\begin{table*}[h]
  \centering
  \caption{An example of assessing drivers' risk-taking behaviour from their driving incidents}
    \begin{tabular}{|l|c|c|c|c|c|c|}
    \hline
    \multirow{3}[6]{*}{} & \multicolumn{4}{c|}{\textbf{Driving incidents}} & \multicolumn{1}{c|}{\multirow{3}[6]{*}{\textbf{ }}} & \multicolumn{1}{c|}{\multirow{3}[6]{*}{\textbf{}}} \\
\cline{2-5}          & \multicolumn{1}{c|}{Frequency  of  } & \multicolumn{1}{c|}{Over-Speed } & \multicolumn{1}{c|}{Excessive Throttle } & \multicolumn{1}{c|}{Frequency  of } &  \textbf{Risk level of }   & \textbf{Risk score(\%)} \\
         & \multicolumn{1}{c|}{Harsh  Braking} & \multicolumn{1}{c|}{duration} & \multicolumn{1}{c|}{ duration} & \multicolumn{1}{c|}{ OverRevving events} &    \textbf{driving behaviour}   &  \\
    \hline
    \textbf{Driver A} & \textcolor[rgb]{ 1,  0,  0}{\textbf{High}} & \textcolor[rgb]{ 1,  0,  0}{\textbf{High}} & \textcolor[rgb]{ 0,  .439,  .753}{\textbf{Moderate}} & \textcolor[rgb]{ 0,  .69,  .314}{\textbf{Low}} & \textcolor[rgb]{ 1,  0,  0}{\textbf{Very High}} & \textcolor[rgb]{ 1,  0,  0}{\textbf{83.09}} \\
    \hline
    \textbf{Driver B} & \textcolor[rgb]{ 1,  0,  0}{\textbf{High}} & \textcolor[rgb]{ 1,  0,  0}{\textbf{High}} & \textcolor[rgb]{ 0,  .439,  .753}{\textbf{Moderate}} & \textcolor[rgb]{ 0,  .69,  .314}{\textbf{Low}} & \textcolor[rgb]{ 1,  0,  0}{\textbf{Very High}} & \textcolor[rgb]{ 1,  0,  0}{\textbf{83.09}} \\
    \hline
    \textbf{Driver C} & \textcolor[rgb]{ 0,  .69,  .314}{\textbf{Low}} &\textcolor[rgb]{ 1,  0,  0}{\textbf{High}} & \textcolor[rgb]{ 1,  0,  0}{\textbf{High}} & \textcolor[rgb]{ 0,  .69,  .314}{\textbf{Low}} &  \textcolor[rgb]{ 1,  0,  0}{\textbf{High}} &  \textcolor[rgb]{ 1,  0,  0}{\textbf{75.24}} \\
    \hline
    \textbf{Driver D} & \textcolor[rgb]{ 0,  .69,  .314}{\textbf{Low}} & \textcolor[rgb]{ 1,  0,  0}{\textbf{High}} & \textcolor[rgb]{ 1,  0,  0}{\textbf{High}} & \textcolor[rgb]{ 0,  .69,  .314}{\textbf{Low}} &  \textcolor[rgb]{ 1,  0,  0}{\textbf{High}} &  \textcolor[rgb]{ 1,  0,  0}{\textbf{75.24}} \\
    \hline
    \end{tabular}%
  \label{tab:riskscore}%
\end{table*}%

\begin{table*}[h]
  \centering
  \caption{An example of moderating drivers' risk-taking behaviour using the effects of contextual factors}
    \begin{tabular}{|c|c|c|c|c|c|}
    \hline
    \multirow{2}[4]{*}{} & \multicolumn{1}{c|}{\multirow{1}[1]{*}{\textbf{Risk score  }}} & \multicolumn{2}{c|}{\textbf{Contextual factors}} & \multicolumn{1}{c|}{\multirow{1}[1]{*}{\textbf{Risk level}}} & \multicolumn{1}{c|}{\multirow{1}[1]{*}{\textbf{Risk score}}} \\
\cline{3-4}       &   \textbf{before moderation (\%) }  & \multicolumn{1}{c|}{Time Pressure} & \multicolumn{1}{c|}{Weather condition} &   \textbf{after moderation}   & \textbf{after moderation (\%)} \\
    \hline
    \textbf{Driver A} & \textcolor[rgb]{ 1,  0,  0}{\textbf{83.09}} & \textcolor[rgb]{ 1,  0,  0}{\textbf{High}} & \textcolor[rgb]{ 1,  0,  0}{\textbf{Rainy}} & \textcolor[rgb]{ 0,  .439,  .753}{\textbf{Moderate}}  & \textcolor[rgb]{ 0,  .439,  .753}{\textbf{63.48}}\\
    \hline
    \textbf{Driver B} &  \textcolor[rgb]{ 1,  0,  0}{\textbf{83.09}} & \textcolor[rgb]{ 0,  .69,  .314}{\textbf{Low}} & \textcolor[rgb]{ 1,  0,  0}{\textbf{Rainy}}  &  \textcolor[rgb]{ 1,  0,  0}{\textbf{Very High}} & \textcolor[rgb]{ 1,  0,  0}{\textbf{90.06}}\\
    \hline
\cline{3-4}       &    & \multicolumn{1}{c|}{Multi-cameras} & \multicolumn{1}{c|}{Emotion} &   & \\
    \hline
    \textbf{Driver C} &  \textcolor[rgb]{ 1,  0,  0}{\textbf{75.24}} &  \textcolor[rgb]{ 1,  0,  0}{\textbf{Absence}} & \textcolor[rgb]{ 0,  .69,  .314}{\textbf{Energetic}}  &  \textcolor[rgb]{ 1,  0,  0}{\textbf{High}} & \textcolor[rgb]{ 1,  0,  0}{\textbf{76.97}}\\
    \hline
    \textbf{Driver D} &  \textcolor[rgb]{ 1,  0,  0}{\textbf{75.24}} & \textcolor[rgb]{ 0,  .69,  .314}{\textbf{Presence}} & \textcolor[rgb]{ 0,  .69,  .314}{\textbf{Energetic}}  & \textcolor[rgb]{ 1,  0,  0}{\textbf{Very High}} & \textcolor[rgb]{ 1,  0,  0}{\textbf{83.59}}\\
    \hline
    \end{tabular}%
  \label{tab:moderation}%
\end{table*}%

In this section, we show the application of the outputs of EDA (GT2 FSs) in moderating and explaining the assessment of driving behaviours using simple examples from the HGV industry as shown in Table~\ref{tab:riskscore}. We replicate a driver risk-scoring system proposed by Mase \textit{et al}~\cite{mase2020capturing}. The driver risk-scoring system uses driving incidents (i.e. number of Harsh Braking (HB) events; Over Speed (OS) duration in seconds; Excessive Throttle (ET) duration in seconds, and number of Over Rev (ORev) events) from HGVs to classify drivers according to the risk of their driving styles on a scale of 0 to 100, where 0 is a low risk driver and 100 high risk. In Table~\ref{tab:riskscore}, we present the risk category and score of four drivers (A, B, C and D) whose driving incidents and scores were simulated using the system. The labels `Low', `Moderate', `High' and `Very High' describe the occurrence of driving incidents and level of risk. For simplicity, we ignore the membership values of the labels by considering only the labels with majority membership. For example, the number of HB events produced by driver A is fuzzified as `Moderate' with a membership value of 0.2 and as `High' with a membership of 0.8; we select `High' due to its higher membership. We use examples with similar risk scores to demonstrate how the impact of external factors could be used to regulate the assessments of driver behaviours. Drivers A and B are classified as `Very High' risk drivers with a risk-score of 83.09\% due to `High' HB and `High' OS incidents while drivers C and D are classified as `High' risk drivers with a risk-score of 75.24\% due to `High' OS and 'High' ET incidents. These assessments produced by the risk-scoring system are based solely on the number of incidents and do not consider contextual factors. 

Table~\ref{tab:moderation} shows regulated driver risk-taking scores when Time Pressure, Multi-cameras, Driver Affective State and Weather conditions are considered. We are assuming Driver A was driving under `High' time pressure while Driver B was under 'Low' time pressure from their managers, and both drivers were driving under the same weather condition. Similarly, Driver C was driving in the 'Absence' of cameras while Driver D in the 'Presence' of cameras, and both drivers were energetic. We use the GT2 FSs obtained for time pressure (see Fig.~\ref{fig:effect_work}), in-vehicle technologies (see Fig.~\ref{fig:effect_tech}), weather conditions (see Fig.\ref{fig:effect_weather}), and affective state (see Fig.\ref{fig:effect_affect}) to understand and regulate the effects of contextual factors on the drivers' risk-taking behaviours as follows:

\begin{enumerate}

\item \textbf{Defuzzify GT2 FSs:} We defuzzify the FSs for high time pressure, low time pressure, absence of cameras, presence of cameras, rainy and energetic using centroid defuzzification~\cite{ginart2002fast} to produce crisp impact scores for the factors, which represent the consensus of the experts' opinions. These crisp scores show how each factor affects a driver's performance from 1 (strong negative impact) to 9 (strong positive impact), and will be utilised to moderate their driving risk scores. The defuzzified impact scores are: high time pressure=2.45, low time pressure=7.55, absence of cameras=4.36, presence of cameras=5.76, rainy=3.78, and energetic=6.02. 

\item \textbf{Merge the impact scores:} To merge the impact scores of the different contextual factors, we employ the commonly used average or mean ensemble method~\cite{wu2003fusing}. We assume joint effects of contextual factors. Other ensemble methods that could be employed are minimum (which prioritise more negative impact factors), weighted average voting (where each factor is assigned a weight), fuzzy rules (where experts determine the combined effects of factors) or Bayes ensemble (where prior information about certain factors are considered)~\cite{wu2003fusing}. The joint effect of the contextual factors for Driver A = mean(2.45, 3.78) = 3.11, Driver B = mean(7.55, 3.78) = 5.67, Driver C = mean(4.36, 6.02) = 5.19, and Driver D = mean(5.76, 6.02) = 5.89.

\item \textbf{Moderate the assessment of driving behaviours:} For simplicity, we could for instance employ a product operator to moderate the assessment of driving behaviours. First, we perform \textit{[x, y]} normalisation of the joint effect, \textit{[x, y]} represents a sensible moderation domain. After several simulations of \textit{x} and \textit{y}, we choose [0.5, 1.5] normalisation for this example as it seems most appropriate to reduce or increase the risk score by half for strong negative or positive influence respectively. It is important to note that \textit{[x, y]} is selected by the decision maker and it controls the moderation magnitude of the decision or outcome. This results to Driver A's contextual effect = $norm(3.11)_{[0.5,1.5]}$ = 0.764, Driver B = $norm(5.67)_{[0.5,1.5]}$ = 1.084, Driver C = $norm(5.19)_{[0.5,1.5]}$ = 1.023, and Driver D = $norm(5.89)_{[0.5,1.5]}$ = 1.111. Lastly, we use the product operator to regulate drivers' risk-taking behaviours as follows: 
\end{enumerate}

\begin{align*}
\text{A's moderated risk score} = 83.09\% \times 0.764 = 63.48\% \\
\text{B's moderated risk score} = 83.09\% \times 1.084 = 90.06\% \\
\text{C's moderated risk score} = 75.24\% \times 1.023 = 76.97\% \\
\text{D's moderated risk score} = 75.24\% \times 1.111 = 83.59\% \\
\end{align*}

The above examples illustrate how the assessment of HGV driving behaviours could be fairly moderated to take into account the driving conditions and external factors. Drivers A and B had the same driving risk score of 83.09\% (very high risk) due to their aggressive driving styles i.e. high HB and OS incidents. Both drivers were driving in the same weather condition (rainy) but under different time pressure to deliver their goods. We observe a significant reduction in the risk-score of driver A after considering the inevitable high time pressure. The results imply that driver A would have been of `Moderate' risk if the driver was rather on `low' time pressure compared to driver B who was under `low' time pressure and still obtained a `Very High' risk score. Similarly, drivers C and D had the same driving risk score of 75.24\% (high risk) due to their speeding driving styles i.e. high OS and ET incidents. Both drivers were driving with the same affective state (energetic) but under different in-vehicle technologies. The moderated scores imply that driver D would have been more risky (risk score of 83.59\%) without the presence of multi-cameras. Therefore, with the impact of contextual factors (GT2 FSs) embedded in intelligent systems, decision makers can clearly understand how different factors may have influenced drivers' behaviours to obtain fairer and more reliable assessments of the drivers.

\section{Conclusion}
\label{conclusion}
This study presents an Expert-centered Driver Assessment (EDA) methodology for capturing the effects of contextual factors on commercial driving, and embedding the opinions of experts into intelligent systems for automatic moderation of driving behaviour assessment. EDA consists of: 1) the identification of contemporary factors, 2) the effective design of data collection tools, 3) the comprehensive collection, modelling and aggregation of expert knowledge about the contemporary factors, and 4) the integration of expert knowledge into intelligent systems. The application of EDA to HGV driving in the UK produce the following results and insights:

\begin{enumerate}
\item A significant number of additional contextual factors are identified by domain experts.

\item There exist variability in the opinions of experts within the same profession indicating the need to recruiting a good number of experts for more comprehensive decisions.

\item There exist variability in the opinions of experts across different professions indicating the need to consider and aggregate knowledge from a wide variety of stakeholders for more reliable and fair decisions.

\item The IAA FSs effectively model interval-valued responses without loss of information and provide clear representations of the variability in experts' opinions within the same professions.

\item The GT2 FSs aggregate experts' opinions from different professions and provide clear visualisations of the degree of agreement among the different professions. 

\item The resulting GT2 FSs can be efficiently embedded into intelligent systems to moderate their outputs and provide interpretable decisions. 

\end{enumerate}

The study shows the importance of collective intelligence of stakeholders in the driving community to identify, understand and embed the influence of personal traits and external contextual factors when evaluating driving behaviours. The all-inclusive approach assists in explaining the decisions of intelligent systems and enables fairness and accountability in the assessment of driving behaviours that lead to incidents or accidents. It is important to note that our sample sizes (the number of participants recruited for the workshops and completing the questionnaires) should be carefully considered in making any statistical conclusions.

For future work, we plan to include additional contextual factors identified during follow-up interviews with experts, such as, road geometry and foggy weather. We also intend to recruit more experts to complete our questionnaire to ensure more robust, comprehensive and statistical results. Lastly, we plan to incorporate the fuzzy sets obtained from this study in the automatic moderation of driver data for fair and explainable assessment of driving behaviours. 

\section*{Acknowledgement}
This work was supported by the Horizon Centre for Doctoral Training at the University of Nottingham and Microlise (UKRI Grant No. EP/L015463/1), and the participants who dedicated some time to complete our questionnaire. We also acknowledge -Lisa Keogh and Neil Selby from Microlise and Shazmin Majid from University of Nottingham- who helped during the study. 

\bibliographystyle{unsrt}
\balance
\bibliography{understandhgv.bib}

\begin{IEEEbiography}[{\includegraphics[width=1in,height=1.15in,clip,keepaspectratio]{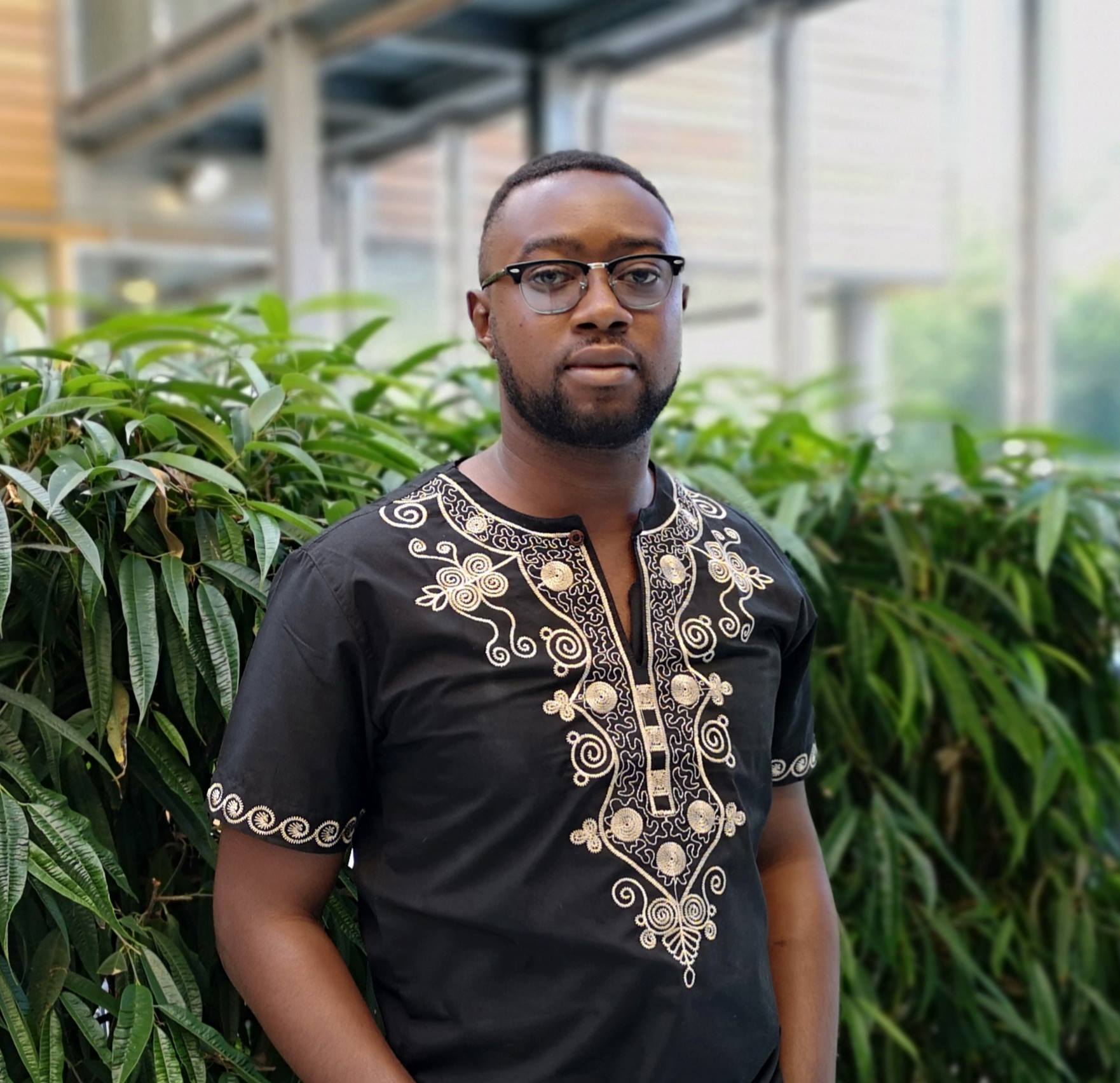}}]{Jimiama Mafeni Mase}

is a PhD candidate in Computer Science at the University of Nottingham. His PhD explores the use of domain experts to fuse and moderate intelligent systems' decisions for a fair and comprehensive assessment of commercial driving behaviours. His expertise includes optimisation, data modelling, information fusion, computer vision, data stream mining, anomaly detection, uncertainty and explainable AI. He is currently a member of Computer Vision Lab (CVL) and the Lab for Uncertainty in Data and Decision Making (LUCID) in the University of Nottingham.  

\end{IEEEbiography}

\begin{IEEEbiography}[{\includegraphics[width=1in,height=1.15in,clip,keepaspectratio]{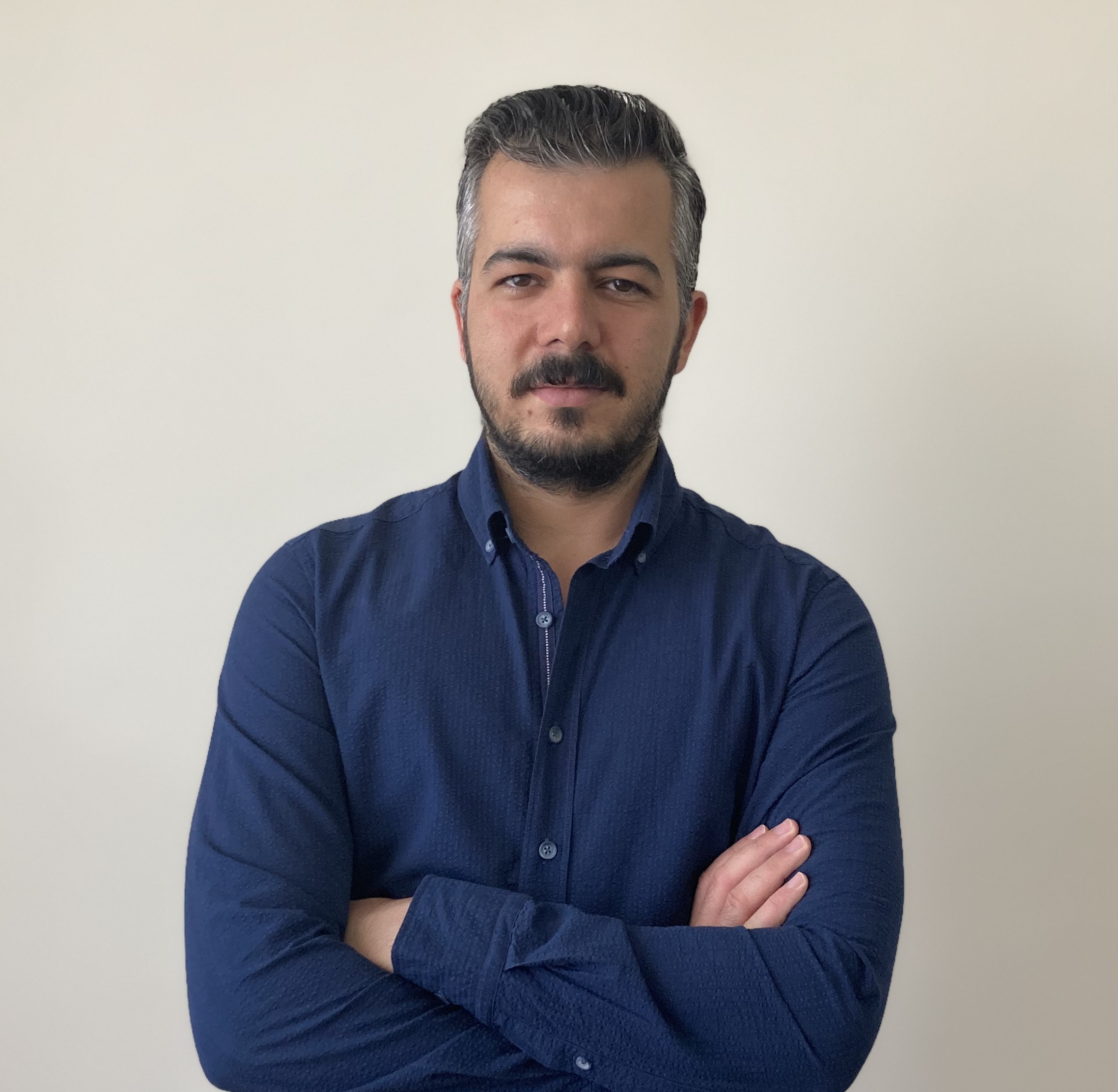}}]{Direnc Pekaslan}

received his MSc degree -with distinction- in the Advanced Computer Science programme from the University of Sheffield, UK., in 2015. He has awarded the PhD degree with the thesis title of 'Towards Better Performance in the Face of Input Uncertainty while Maintaining Interpretability in AI' at the School of Computer Science, University of Nottingham, UK., in 2021. He is currently a research fellow in the project 'Leveraging the Multi-Stakeholder Nature of Cyber Security' on human centred cybersecurity studies that explores novel approaches of capturing and modelling uncertain data on the vulnerability of computer systems from a variety of sources. He is currently a member of the Lab for Uncertainty in Data and Decision Making (LUCID) and the Intelligent Modelling and Analysis (IMA) research group. His main research interest includes uncertainty handling with interpretable AI systems, particularly, focuses on providing adaptive behaviours in non-singleton fuzzy models that can handle unexpected circumstances in real-world applications.
\end{IEEEbiography}

\begin{IEEEbiography}[{\includegraphics[width=1in,height=1.15in,clip,keepaspectratio]{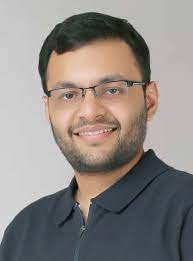}}]{Dr Utkarsh Agrawal}

received his PhD in Computer Science from the University of Nottingham. He is currently a post-doctoral research fellow at University of St Andrews. His research interests include the application of data analytics and machine learning in real-world scenarios and his current focus is on vaccine effectiveness and multimorbidity.

\end{IEEEbiography}

\begin{IEEEbiography}[{\includegraphics[width=1in,height=1.15in,clip,keepaspectratio]{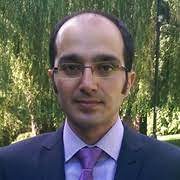}}]{Dr Mohammad Mesgarpour}

received the PhD degree in operational research from the University of
Southampton in 2012. He is now head of Data Science and Operational Research in Microlise. Prior to that, he was a KTP Research Associate with The University of Nottingham for two years. His main areas of
research is in the fields of transport management, predictive modelling, data analytics, and combinatorial optimisation.

\end{IEEEbiography}

\begin{IEEEbiography}[{\includegraphics[width=1in,height=1.15in,clip,keepaspectratio]{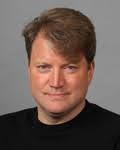}}]{Dr Peter Chapman}

is an Associate Professor, researching applied cognitive psychology within the School of Psychology at the University of Nottingham. His main areas of research are the psychology of driving, the understanding of crashes and their causes (memory, attention, distraction, human error etc.), visual search in novice and experienced drivers, ways of training newly qualified drivers to use more effective visual search strategies, and eye movements in dangerous driving situations. 

\end{IEEEbiography}

\begin{IEEEbiography}[{\includegraphics[width=1in,height=1.15in,clip,keepaspectratio]{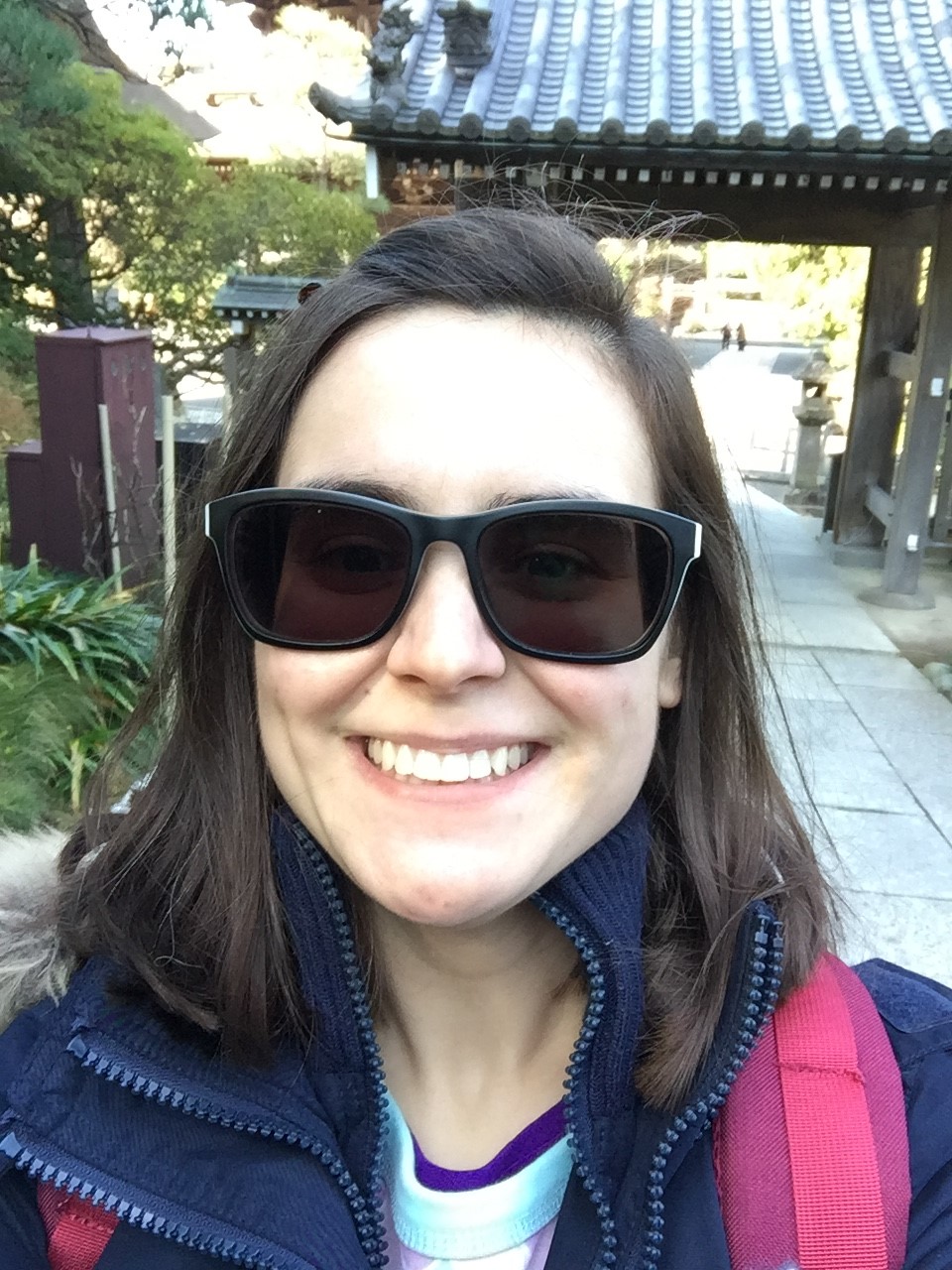}}]{Dr Mercedes Torres Torres}

is an Assistant Professor of Computer Science. She is a member of the Computer Vision Lab (CVL) and the Horizon Research Institute. Her area of expertise is in machine learning and computer vision, with particular emphasis in the development of novel deep learning techniques for small and skewed datasets. Her research portfolio includes multidisciplinary collaborations in areas like Healthcare (QMC), GIS (Dstl), Cybersecurity (Netacea), and Children Protection Services (Nottingham City Council), with funding from EPSRC (Horizon Research Institute, TAS Network, EnLightenUs as Co- I), Innovate UK (KTP, PI), and Huawei (HIRP, PI). She has published in high-impact conferences and journals, such as LREC, AIME, ESWA, FG and IMAVIS.

\end{IEEEbiography}

\begin{IEEEbiography}[{\includegraphics[width=1in,height=1.15in,clip,keepaspectratio]{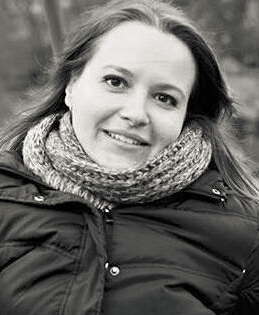}}]{Dr Grazziela Figueredo}

is Assistant Professor in the School of Computer Science and the Research and Innovation Lead at the Digital Research Service at The University of Nottingham. The focus of her research is the development and application of techniques for systems simulation and intelligent data analysis for urban mobility and transportation. Grazziela has served as Associate Editor of the IEEE Intelligent Transportation Systems conference and has published over 20 papers in AI for transport research (in journals such as IEEE Transactions on Intelligent Transportation Systems and Accident Analysis and Prevention).

\end{IEEEbiography}

\onecolumn
\appendix

\begin{figure}[ht!]
\centering
\scalebox{0.25}{
\includegraphics{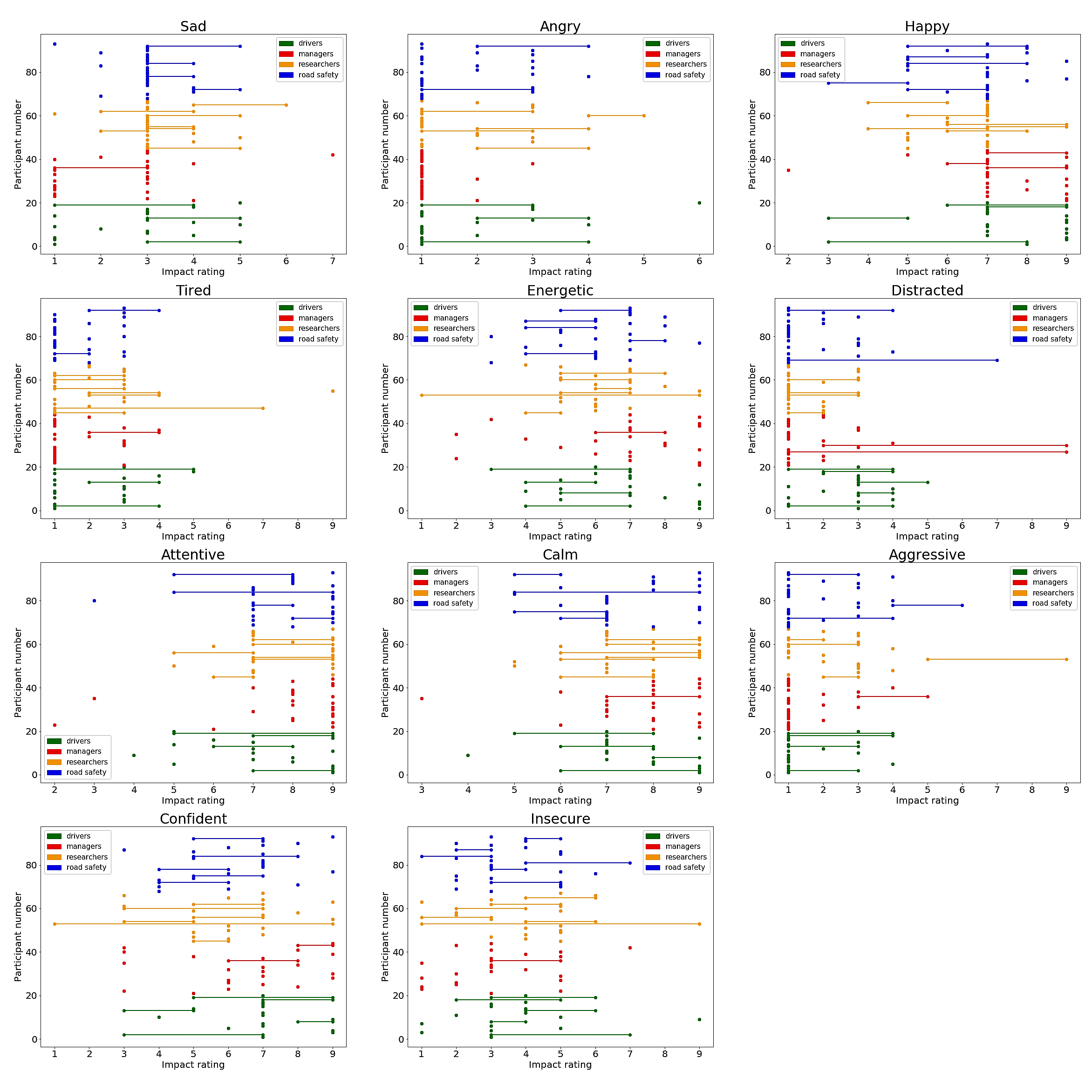}
}
\caption{line graphs showing the individual responses from experts about the impact of HGV drivers' affective states on their driving performance.}

\label{fig:affect_line}
\end{figure}


\begin{figure}[ht!]
\centering
\scalebox{0.25}{
\includegraphics{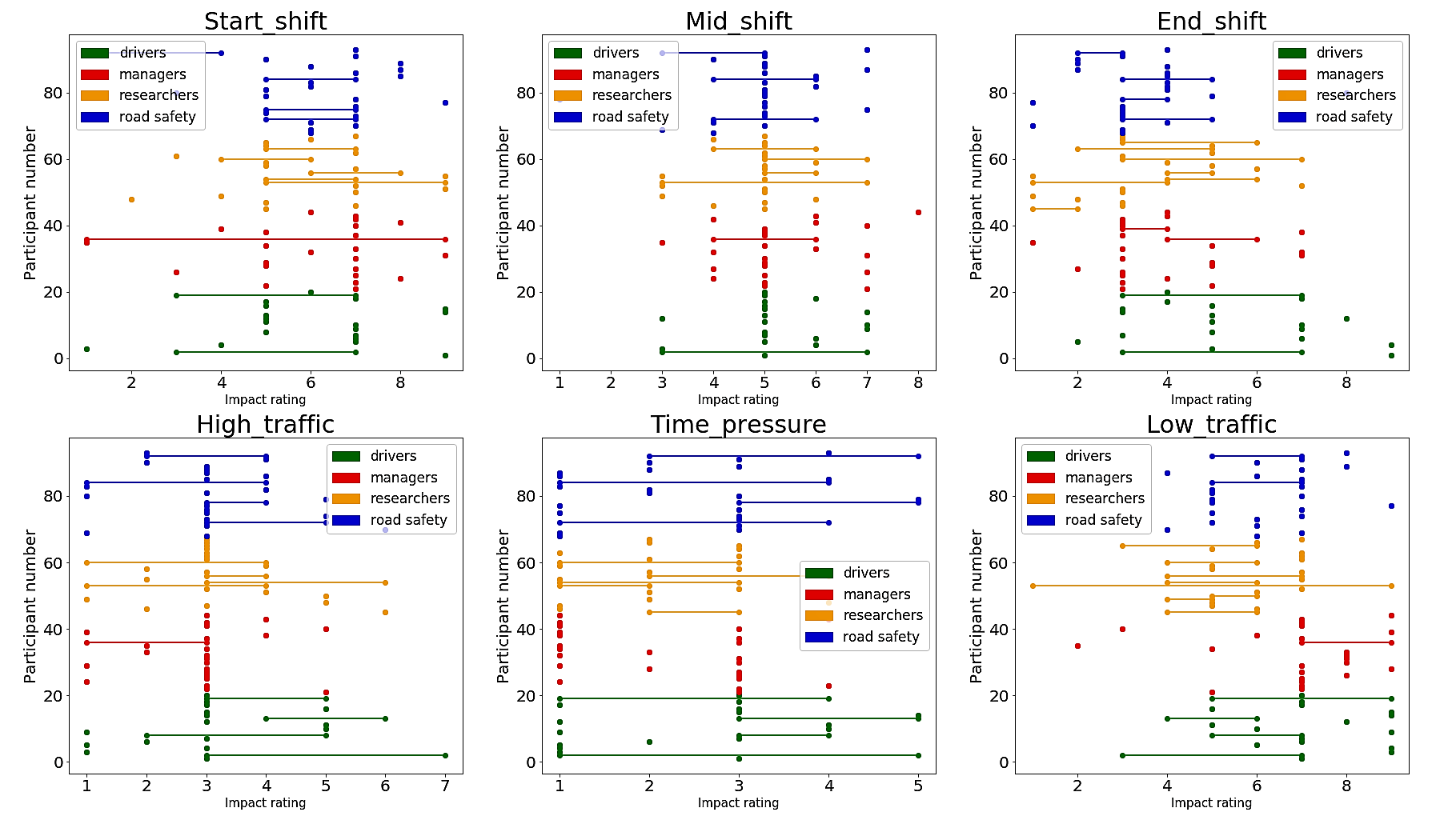}
}
\caption{line graphs showing the individual responses from experts about the impact of work related factors on HGV driving performance.}

\label{fig:work_line}
\end{figure}


\begin{figure}[ht!]
\centering
\scalebox{0.25}{
\includegraphics{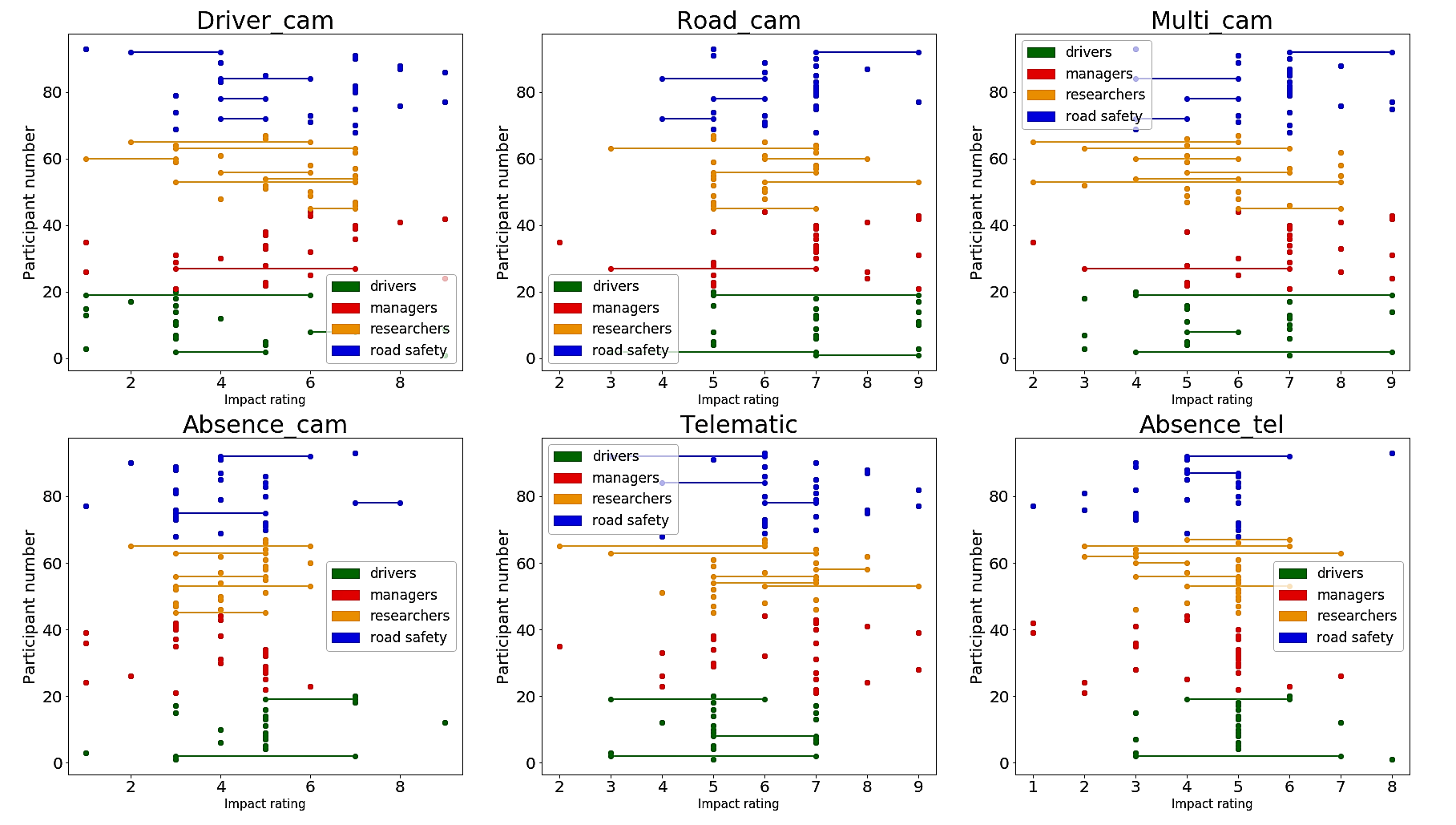}
}
\caption{line graphs showing the individual responses from experts about the impact of technologies on HGV driving performance.}

\label{fig:work_line}
\end{figure}


\begin{figure}[ht!]
\centering
\scalebox{0.25}{
\includegraphics{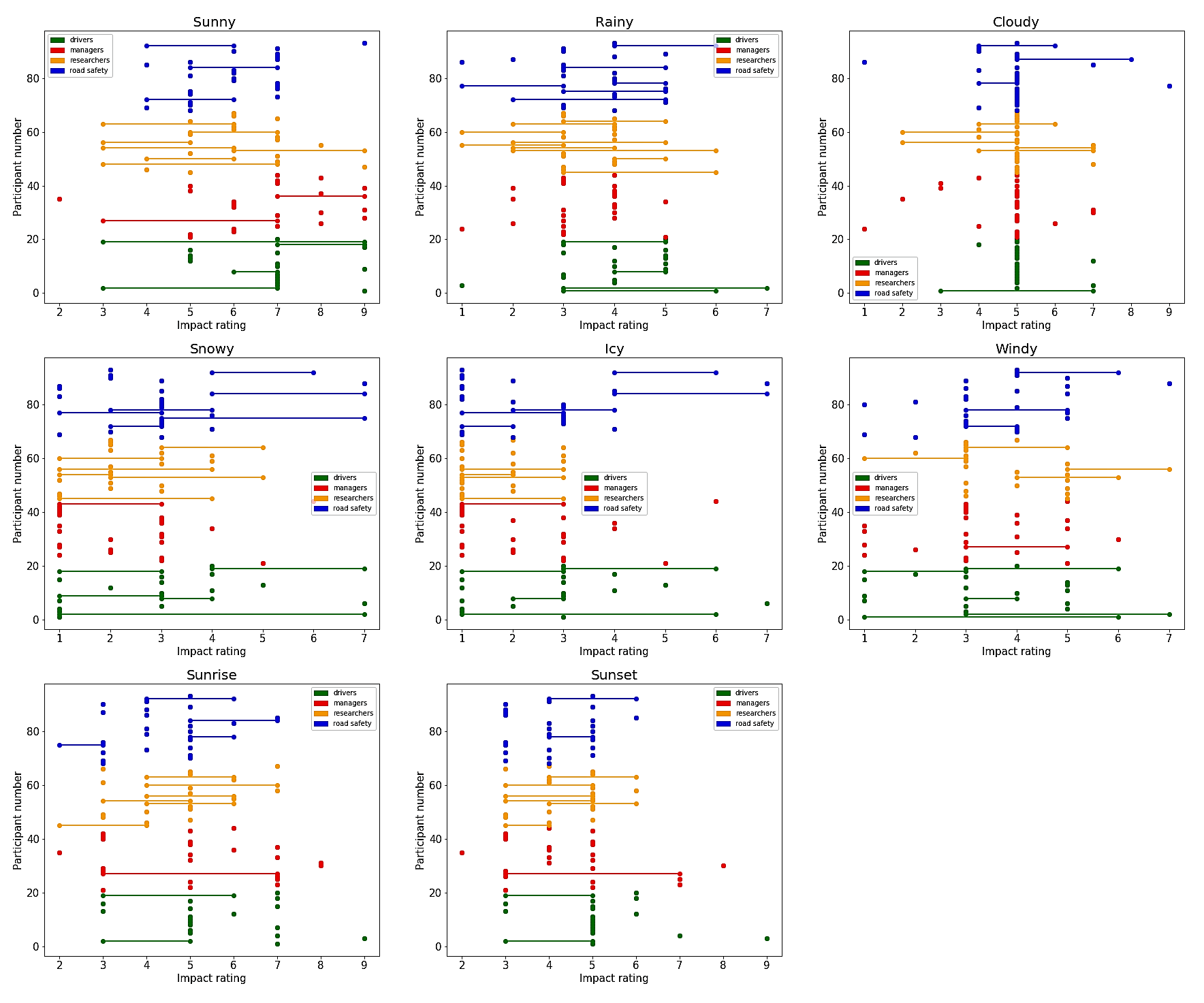}
}
\caption{line graphs showing the individual responses from experts about the impact of weather conditions on HGV driving performance.}

\label{fig:work_line}
\end{figure}

\end{document}